\pdfoutput=1
\PassOptionsToPackage{dvipsnames}{xcolor}

\documentclass[11pt]{article}

\usepackage[final]{acl}

\usepackage{times}
\usepackage{latexsym}
\usepackage{longtable}
\usepackage{rotating}
\usepackage{pdflscape}
\usepackage[T1]{fontenc}

\usepackage[utf8]{inputenc}

\usepackage{microtype}

\usepackage{inconsolata}

\usepackage{graphicx}
\usepackage{enumitem}
\usepackage{comment}
\usepackage{amsmath}
\usepackage{amsfonts}
\usepackage{gb4e}
\noautomath
\usepackage{booktabs} 
\usepackage{dashrule}
\usepackage{multirow} 
\usepackage{tabularx}
\usepackage{caption} 
\usepackage{listings}
\usepackage{soulpos}
\usepackage{subcaption}
\usepackage{tcolorbox}
\tcbuselibrary{breakable}
\tcbset{
  colframe=black,          
  colback=white,           
  breakable,               
}

\newcommand{\naoki}[1]{#1}

\newcommand{\sgn}{\operatorname{sgn}}

\newcommand{\system}{\textsc{SnipRec}}

\newcommand{\snippet}[2][blue]{#2}

\title{Snippet-based Conversational Recommender System}

\author{Haibo Sun \thanks{This work was conducted while Haibo Sun was a research intern at Megagon Labs.}\\
  Brandeis University \\
  \texttt{hsun@brandeis.edu} \\\And
  Naoki Otani \\
  Megagon Labs \\
  \texttt{naoki@megagon.ai} \\\And
  Hannah Kim \\
  Megagon Labs \\
  \texttt{hannah@megagon.ai} \\\AND
  Dan Zhang \\
  Megagon Labs \\
  \texttt{dan\_z@megagon.ai}\\\And
  Nikita Bhutani \\
  Megagon Labs \\
  \texttt{nikita@megagon.ai}}

\begin{document}
\maketitle

\begin{abstract}

Conversational Recommender Systems (CRS) engage users in interactive dialogues to gather preferences and provide personalized recommendations. While existing studies have advanced conversational strategies, they often rely on predefined attributes or expensive, domain-specific annotated datasets, which limits their flexibility in handling diverse user preferences and adaptability across domains. We propose \system{}, a novel resource-efficient approach that leverages user-generated content, such as customer reviews, to capture a broader range of user expressions. By employing large language models to map reviews and user responses into concise \textit{snippets}, \system{} represents user preferences and retrieves relevant items without the need for intensive manual data collection or fine-tuning. Experiments across the restaurant, book, and clothing domains show that snippet-based representations outperform document- and sentence-based representations, achieving Hits@10 of 0.25-0.55 with 3,000 to 10,000 candidate items while successfully handling free-form user responses.

\end{abstract}

\section{Introduction}

Conversational Recommender Systems (CRS) aim to gather user preferences through conversation and provide personalized recommendations based on user responses. This approach is particularly useful when users lack domain knowledge, are unsure of their preferences, or when the system has no prior user information (the \textit{cold-start} problem). To achieve this, a CRS must effectively: a) organize information about target items, b) interpret user responses to identify relevant items, and c) pinpoint significant aspects to further solicit user preferences. This paper focuses on tasks (a) and (b), which have received less attention than (c). We propose a method that uses large language models (LLMs) to handle the wide variety of domain-specific topics naturally arising in conversations.

\begin{figure}
    \centering
    \includegraphics[width=1.0\linewidth]{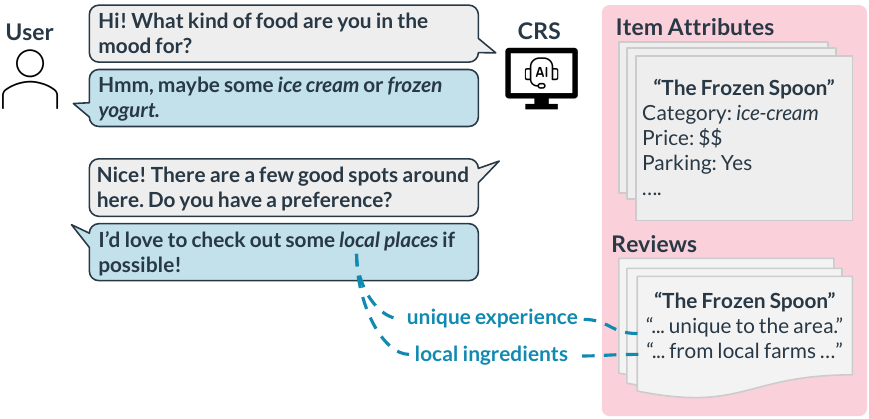}
    \caption{Example interactions with a conversational recommender system (CRS) driven by customer reviews. CRS informed by customer reviews can manage diverse user expressions that cannot be handled solely by item attributes and more effectively retrieve relevant items.}
    \label{fig:intro}
\end{figure}
Data-driven strategies for (c) conversation strategies have seen significant progress through extensive research (see, e.g., \citealp{sun2018conversational,lei2020estimation,lei2020interactive,deng2021unified}). The majority of these studies depend on predefined item attributes, such as restaurant categories or movie genres, assuming that (a) all items fall into these categories and (b) user responses can be mapped onto these attributes and their values. Another major line of work has developed CRS systems that mimic human conversation strategies (e.g., \citealp{li2018towards,kang2019recommendation}). In this approach, flexible, human-like conversations result from learning implicit representations of (a) items and (b) user responses based on large conversation datasets.

Yet, collecting conversation data that covers a wide variety of items and user preferences remains challenging, and adapting to new domains is difficult. In this paper, we aim to advance CRS by embracing a wide range of user expressions and preferences and enabling easy adaptation to new domains without additional in-domain data collection or fine-tuning. We focus on the efficient acquisition of item information and the flexible interpretation of user utterances, a direction orthogonal to conversation strategies. Specifically, we utilize user-generated content (UGC) like customer reviews to capture domain-specific and diverse user expressions and preferences. Systems supported by rich UGC data can potentially manage any information users mention and connect them to relevant items more effectively (illustrated in Fig. \ref{fig:intro}). We leverage LLMs for their strengths in language understanding and information extraction \cite{wei2023zero, li2023evaluating} to derive high-quality insights from UGC, improving conversational recommendations in a resource-efficient and adaptable manner.

We introduce \system{}, a system that uses \textbf{snippets} mined from UGC~(\S\ref{sec:sniprec}). A snippet is an atomic unit of information that conveys either an objective fact or a subjective opinion and serves to represent items and capture user preferences. Their translation to snippets is executed by LLMs in a few-shot manner without the need for domain-specific data annotation or fine-tuning. We offload the conversation strategy of SnipRec to LLMs, which have proven effective in eliciting information from users \cite{li2023eliciting,he2023large}.\footnote{The conversation module can be replaced with other advanced methods, but that is beyond the scope of this paper.}

To evaluate \system{} at scale, we employ an LLM-based user simulation--a new paradigm for CRS evaluation \cite{yoon2024evaluating, wang2023rethinking}. The simulator models a diverse set of users using selected items and reviews, allowing conversations on a wide range of topics~(\S\ref{sec:user-simulator}). Experiments on three datasets from distinct domains (restaurants, books, and clothing) show that snippet-based representations consistently improve item retrieval performance compared to document and sentence-based representations in CRS~(\S\ref{sec:experiments}). In particular, \system{} using both GPT-4o-mini and LLaMA-3.3-70B models improved Hits@10 by 0.1-0.25. We also observed that the LLM-based snippet extraction and user simulation methods performed reliably in most cases (>97\% and >90\% of faithfulness to the context, respectively) through automatic and manual evaluations.

This study advances previous CRS work in several ways. (1) We leverage UGC to capture a wider range of diverse expressions and long-tail information beyond predefined attributes. (2) We introduce \system{}, an LLM-driven CRS approach that represents item information and user preferences using snippets. While achieving the aforementioned goals, our method reduces the need for domain-specific annotation or training, allowing straightforward domain adaptation. (3) We extend a recent LLM-based user simulator to reliably evaluate multi-turn CRS involving free-form conversations.

\section{Task and Problem Setting}

Given a set of candidate items (e.g., restaurants, hotels) with user-generated content (e.g., reviews), the goal is to recommend the most suitable item through a multi-turn conversation between a \textit{seeker} (user) and a \textit{recommender} (system).
The recommender focuses on two primary tasks: search and conversation strategy.

\begin{itemize}[leftmargin=10pt]
\setlength{\itemsep}{-0.3em}
     \item \textbf{Search} aims to retrieve relevant items based on the implicit and explicit preferences of the seeker. At each turn, the seeker responds to a question from the recommender. The recommender then uses this information to retrieve and rank candidate items.
    \item \textbf{Conversation strategy} aims to ask clarifying questions for multiple turns to elicit specific preference information from the seeker, which helps to refine the search space effectively.
   
\end{itemize}
These two tasks work together, with the search refining the recommendation pool and the conversation strategy uncovering valuable preferences. This iterative approach ensures a dynamic and user-centric dialogue experience that adapts to the user's needs. We assume that the seeker has a specific target item in mind and only provides relevant information to the open questions asked by the system without offering extraneous information. The overall objective is to minimize the number of turns needed to identify the seeker's target item.

\section{Methodology}
\label{sec:sniprec}

\begin{figure}
    \centering
    \includegraphics[width=0.9\linewidth]{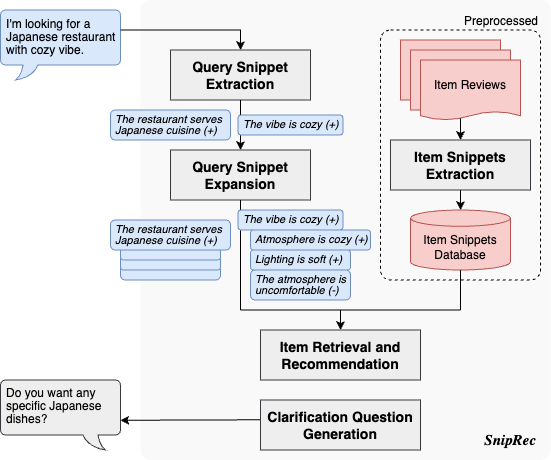}
    \caption{\system ~workflow. 
    }
    \label{fig:flow}
\end{figure}

Fig.~\ref{fig:flow} provides an overview of \system{}. Item reviews are first extracted into \textit{item} snippets to represent item knowledge (\S\ref{sec:snippet_review}). During the conversation, seeker's responses are parsed into \textit{query} snippets (\S\ref{sec:snippet_user}). The recommender uses these query snippets to retrieve relevant item snippets and rank items (\S\ref{sec:retrieval}). If the target item is not ranked at the top, it generates a clarification question based on the conversational history~(\S\ref{sec:cq}).

\subsection{Snippets from Item Reviews}
\label{sec:snippet_review}

People often express their needs in various ways~\cite{lyu2021workflow}, such as ``I am looking for a restaurant for a family gathering.'' and ``I'd prefer a place with a good view.'' These free-form expressions can go beyond few predetermined item attributes which many existing CRS studies rely on. In contrast, UGC like customer reviews captures any information that matters to users. Reviews offer opinions and facts about various aspects of items, making them a rich source for mining item knowledge.

Reviews may be used as a complete document~\cite{wang2023rethinking} or segmented by sentence~\cite{lu2021revcore}, but its nature poses challenges. Reviews can be verbose or omit details, which calls for more flexible information extraction. To obtain useful insights, we propose decomposing reviews into short snippets, each conveying atomic meaning~\cite{nenkova2004evaluating}, inspired by recent advances in fact-checking~\cite{wanner2024closer, min2023factscore}.

Motivated by the strong language understanding and information extraction capabilities of LLMs, we use LLMs with a few-shot learning method to decompose item reviews into \snippet[red]{\textbf{item}} snippets, focusing on factual descriptions or subjective opinions related to a specific aspect of the item~(The prompt can be found in Appendix~\ref{sec:prompt_item_snippet}). Specifically, given an item $i$ and its associated reviews $R^i$, we extract item snippets from each review and merge them into this item's snippet set, denoted as $S^i=\bigcup\limits_{r\in R^i}\mathsf{DECOMP}(r)$. As shown in the example below, snippet extraction goes beyond simple parsing:
\vspace{-0.25\baselineskip}
\begin{description}
\setlength{\itemsep}{1pt}
\setlength{\parskip}{0pt}
    \item[$i$ = ] The Kebab Kitchen (restaurant)
    \item[$R^i$ =] {"Perfect for a family gathering party...", "We ordered classic hummus, chicken kebab.", ...}
    \item[$S^i$ =] "This restaurant is perfect for a family gathering party.", "This restaurant serves classic hummus.", "This restaurant serves chicken kebabs.", ...
\end{description}
This process often necessitates flexible filtering, completion, and rephrasing, a task at which LLMs excel.\footnote{While LLMs may introduce hallucination (instances where extracted snippets contain non-existent information from the original review text), our evaluation shows that such errors are quite rare~(\S\ref{sec:snippet_evaluation}). In general, the snippets are finer-grained than the original text.} Additionally, we transform the predefined attributes in the dataset for $i$ into snippets using templated sentences as described in Appendix \ref{sec:attribute_snippet}.

\subsection{Snippets from Seeker Responses}
\label{sec:snippet_user}

At each turn $t$, the seeker articulates preferences in their response $a_t$. We apply a similar process to extract snippets from $a_t$ (details in Appendix~\ref{sec:prompt_query_snippet}). Specifically, we decompose each user response $a_t$ into several \snippet[blue]{\textbf{query} snippets} by $\mathsf{DECOMP}(a_t)$, using a weakly canonicalized format similar to item snippets. For example, the utterance ``I'm looking for a cozy cafe'' is converted to 
\snippet[blue]{``this cafe has a cozy atmosphere.''} The conversion reduces semantic differences between query and item snippets to support accurate retrieval.

Every query snippet also includes a \textit{sentiment} indicator. This indicator denotes the user's preferences or dislikes for a particular aspect. For instance, if the user says ``I’d love some cake, but I want to avoid noisy places,'' the decomposition yields two snippets: \snippet[blue]{``this restaurant serves pizza''} and \snippet[blue]{``this restaurant is noisy.''} These snippets are labeled as ``prefer'' and ``dislike'', respectively. 

\paragraph{Snippet Expansion}

Query snippets often express indirect, long-tail subjective preferences with diverse lexical choices. To alleviate retrieval inaccuracies caused by this property, we use LLMs (Appendix~\ref{sec:prompt_query_expansion}) and expand the query snippets through three transformations: \textit{paraphrase}, \textit{support}, and \textit{opposite} inspired by \cite{xu2024personalized, lyu2024llm}. After expansion, the query snippets at turn $t$ are defined as:
\begin{align*}
S^u_t&=\{s, ~\mathsf{paraphrase}(s), ~\mathsf{support}(s), \\
&~~~~~~~\mathsf{opposite}(s):\forall s\in \mathsf{DECOMP}(a_t)\}
\end{align*}
These transformations are exemplified below:
\begin{exe}
\ex $\mathsf{paraphrase}$(\snippet[blue]{``...warm service...''})\\ $\xrightarrow{\text{expand}}$ \snippet[blue]{``...friendly server...''}
\ex $\mathsf{support}$(\snippet[blue]{``...comfortable seating...''})\\ $\xrightarrow{\text{expand}}$ \snippet[blue]{``...cozy atmosphere...''}
\ex $\mathsf{opposite}$(\snippet[blue]{``service is prompt''} (prefer))\\ $\xrightarrow{\text{expand}}$ \snippet[blue]{``wait time is long''} (dislike)
\end{exe}

\subsection{Item Retrieval and Reranking}
\label{sec:retrieval}

Given the query snippets at each turn $S^u_t$, we use dense retrieval \cite{karpukhin2020dense} to identify relevant item snippets from the entire set $S^I$ ($=\cup_{i\in I} S^i$). To ensure the retrieved snippets contain supporting information for each query snippet, we re-rank them using a natural language inference (NLI) model. Then, we score items based on the retrieved and re-ranked snippets. 

\paragraph{Retrieval of Item Snippets}
Concretely, we use an off-the-shelf pretrained encoder to embed item snippets $S^I$ and query snippets $S^u_t$. For each query snippet in $S^u_t$, we retrieve top-$k$ similar item snippets based on cosine similarity, denoted as:

\begin{equation*}
S^{I,u}_t=\bigcup\limits_{s^u_t \in S^u_t}
\mathop{\text{arg~top}~k}\limits_{s^i \in S^I}
~\Bigl(\mathsf{similarity}(\varepsilon(s^i),\varepsilon(s^u_t))\Bigr),
\end{equation*}
where $\varepsilon$ is an embedding function and $|S^{I,u}_t|=|S^u_t|\times k$.\footnote{$k$ is a hyperparameter.}

Next, we rank the item snippets within each group, where the groups are organized based on the query snippets used as the retrieval keys. Since vector search returns semantically similar content but does not guarantee relevance of retrieved item snippets, we use an NLI model to ensure these snippets truly satisfy the query snippets. 


The ranking of a specific item snippet is based on the entailment score between each item snippet and its corresponding query snippet, as provided by the NLI model, and is performed independently within each set of item snippets retrieved by each query snippets, rather than across all retrieved item snippets in one turn.
The higher entailment score, the higher their rank is (thus lower value in ranking position). We discard item snippets with scores below a threshold \(t_\text{entailment}\). To summarize, for each query snippet $s^u_{t,k} \in S^u_t$ (the query snippets extracted from the utterance in the turn $t$), we obtain the rank of an item snippet $s^i \in S^{I, u}_t$ as follows: \begin{align*}
    & \mathsf{rank}_{s^u_{t,k}}(s^i) \propto 1 / \mathsf{NLI}(s^i \xrightarrow{entail} s^u_{t,k})\\
    & \text{s.t.} \quad \mathsf{NLI}(s^i \xrightarrow{entail} s^u_{t,k}) \geq t_{\text{entailment}} 
\end{align*} This process results in a smaller set of item snippets for each query snippet, which we denote as $S'^{I,u}_t$.





\paragraph{Item Re-ranking}
\label{itemranking}

For each item $i$, we now update its score using the retrieved item snippets associated with item it. Their within-group ranks are aggregated to calculate the item score. We employ Reciprocal Rank Fusion~\cite{cormack2009reciprocal}, weighing each item snippet by $1/(\kappa + \mathsf{rank})$.\footnote{$\kappa$ is a hyperparameter. Following common practices, we set $\kappa=60$.}  If multiple snippets for item $i$ are included in the same ranking, we only consider the one with the highest rank, ignoring lower-ranked snippets. The score for item $i$ is calculated as follows:

\vspace{-10pt}
\begin{align*}
\mathsf{SCORE}_t(i) &= \mathsf{SCORE}_{t-1}(i) \\
&+ \sum\limits_{s^u_{t,k}\in S^{u}_t}  \sum\limits_{s^i \in S'^{I, u}_{t} \cap S^{i}} \frac{\sgn(s^u_{t,k})}{\kappa+\mathsf{rank}_{s^{u}_{t,k}}(s^i)}    
\end{align*} where $\sgn$ indicates the sentiment polarity that each query snippet $s^u_t \in S^u_{t,k}$ has: \begin{align*}
    \sgn(s^u_{t,k}) = 
\begin{cases} 
1, & \text{if sentiment is \textit{prefer}} \\
-1, & \text{if sentiment is \textit{dislike}}
\end{cases}
\end{align*} This determines whether the item snippets it retrieves contribute positively or negatively to $\mathsf{SCORE}_t(i)$. A higher positive score results in a higher ranking for the item. Items that do not have any retrieved item snippets are disregarded in the item reranking process.

\subsection{Clarification Questions}
\label{sec:cq}

At each step, the recommender asks a clarification question to capture the seeker's preferences and narrow the search space. As the primary focus of this study is not on conversation strategies, we delegate this step to an LLM, which has shown effectiveness in generating context-specific questions in dialogue settings in recent studies~\cite{li2023eliciting,he2023large}. Although other conversation strategies exist~(\citealp{lei2020estimation,lei2020interactive,xu2021adapting}; \textit{inter alia}), this LLM-based approach meets our need to elicit free-form utterances from users in a resource-efficient way.

We prompt the LLM to produce a relevant question based on the conversation history (Appendix~\ref{sec:prompt_CQ}). This approach avoids extensive data annotation and fine-tuning, facilitating domain adaptation. Note that, similar to existing work \cite{wang2023rethinking}, our question generator depends solely on the LLM’s internal knowledge without incorporating signals from snippet extraction or item retrieval.\footnote{While not central to our main research claim, we discuss a potential method to link snippet retrieval and question generation in Appendix \ref{sec:sampled}.}

\section{User Simulation for Evaluation}
\label{sec:user-simulator}

A user simulator in a CRS imitates real user behavior and preferences during interactions, enabling testing and evaluation without actual user involvement. Recently, LLM-based user simulators have emerged as a promising new approach for CRS evaluation~\cite{yoon2024evaluating,liang2024llm,kim2024pearl}, allowing the scalable simulation of human-like interactions. To address potential challenges associated with LLM limitations, like hallucinations, domain constraints, and answer leakage, we extend the existing framework according to the following guidelines:

\begin{itemize}[leftmargin=10pt]
\setlength\itemsep{-0.3em}
    \item \textbf{Avoid answer leakage.} Instructing LLMs not to disclose answers in contexts cannot prevent cheating~\cite{wang2023rethinking}. We must ensure that the context entirely excludes the answer.
    \item \textbf{Communicate like human.} Unlike humans, LLM responses are often verbose and formal. Controlling verbosity and tone is essential.
    \item \textbf{Stay relevant.} User simulators' answers must fully address the system's question without omitting relevant details.
    \item \textbf{Stay on-topic.} User simulators' answers should be limited to the topic given by the system, without adding extraneous information. 
    \item \textbf{Be faithful to context.} User simulators' answers must remain consistent with the provided context, avoiding contradictions or unsupported opinions.
\end{itemize}

Following these guidelines, we simulate a user searching for a pre-selected, highly rated item from the dataset. To construct a user profile that favors the selected item, we create context using a sampled positive review of the item, a summary of this item's other favorable reviews, and essential attribute information provided in the dataset (e.g., cuisine type and price range for restaurants, or book description). This approach ensures the LLM-based simulator remains focused on both objective and subjective details about the item. We instruct the LLM to stay relevant, on-topic and faithful to the context, aiming to emulate a human's ability to recall preferences during a conversation. The evaluation of our user simulator is presented in Section~\ref{sec:user-simulator-evaluation},  with additional details regarding the prompt provided in Appendix~\ref{sec:prompt_simulator}.

The user simulator generates only natural conversation utterances by providing relevant information, allowing its application across various types of CRS. It is not specifically designed for \system{} and thus it does not favor our proposed method.

\section{Experiments}
\label{sec:experiments}


\naoki{To validate the effectiveness of snippet-level representations for conversational recommendation, we evaluate \system{} on three datasets. We also validate the role of LLMs as conversation managers, evaluate the quality of extracted snippets, and test the reliability of our LLM-based user simulator via manual and automatic analyses.}

\subsection{Experimental Setup}
\label{sec:exp-setting}
\paragraph{Dataset}
We use the Yelp dataset (restaurant)\footnote{https://www.yelp.com/dataset} and Amazon Reviews dataset (book and clothing) \cite{hou2024bridging} which includes reviews and corresponding item information. For Yelp dataset, we extract businesses located in Philadelphia, the city with the most registered businesses in the dataset, and retain only those with ``food'' in their Yelp categories\footnote{This setup simulates real user behavior by concentrating on specific areas and categories.} and at least 10 reviews. For Amazon Reviews dataset, we picked the Books and Clothing categories and sampled 10,000 items with at least 10 reviews, each of which is a verified purchase and has at least 1 helpful vote. Table~\ref{tab:dataset_statistics} shows the statistics of our datasets.

\begin{table}[t]
    \centering
    \small
    \begin{tabular}{lrrrr}
         \toprule
         \textbf{Dataset} & Items & Users & Reviews & Snippets\\
         \midrule
         Restaurant & 3,007 & 129,503 & 170,996 & 986,819 \\
         Book & 10,000 & 232,379 & 249,596 & 999,931 \\
         Clothing & 10,000 & 417,501 & 442,295 & 1,725,964 \\
         \bottomrule
    \end{tabular}
    \caption{Dataset statistics.}
    \label{tab:dataset_statistics}
\end{table}

\paragraph{User Simulator} 
Our LLM-based simulator~(\S\ref{sec:user-simulator}) imitates a seeker with a specific target venue in mind. To do this effectively, we provide the simulator with the following information: a) Yelp categories and attributes of the target item, or Amazon dataset's features, description, authors, categories and prices; b) a summary of positive reviews about the target and c) a review representing seeker's preference, along with general instructions and conversation history. GPT-4o-mini\footnote{We used \texttt{gpt-4o-mini-2024-07-18} for all experiments involving GPT-4o-mini.} serve as the simulator's backbone throughout all experiments.

We use positive reviews as seed data, selecting reliable users and high-quality reviews based on the following criteria: a) review ratings of four or five, b) at least one useful~(Yelp) / helpful~(Amazon) vote for the review, and c) users who have written 10 and 99 reviews. We then identify (user, item) pairs through maximum bipartite matching\footnote{\url{https://docs.scipy.org/doc/scipy/reference/generated/scipy.sparse.csgraph.maximum_bipartite_matching.html}} to maximize pair extraction without overlap, sampling 500 pairs for validation and 1,000 for testing for each dataset. We refer to these selected users as \textit{seed users}. Next, we use GPT-4o-mini for post-processing. Following \citet{kim2024pearl}, we summarize general opinions from non-seed users. For each item, we provide GPT with Yelp/Amazon categories, Yelp attributes or Amazon features \& descriptions, and five positive and five negative reviews (prioritizing those with higher usefulness scores) to generate summaries of five sentences each for positive and negative aspects. Finally, we anonymize proper nouns in both seed reviews and summaries to prevent answer leakage. Note that the seed users' reviews are not used by recommender systems for preventing shortcuts in item retrieval.

\paragraph{Baselines}
To test the effectiveness of the snippet-based representation and the proposed techniques, we compare \system{} with LLM-based systems without query snippet extraction or expansion~(\S\ref{sec:snippet_user}), using three different representations of UGC: (1) a \textbf{document-based} baseline that directly uses raw reviews without decomposition,\footnote{Like our baseline, some studies use review documents without decomposition as part of retrieval targets. (e.g., \citealp{wang2023rethinking})} (2) a \textbf{sentence-based} baseline that splits review documents into individual sentences using  spaCy~\cite{honnibal2020spacy}\footnote{\texttt{en\_core\_web\_sm} was used.}, and (3) a \textbf{snippet-based} baseline that uses the extracted item snippets~(\S\ref{sec:snippet_review}). While sentences are similar to snippets, they tend to be more diverse in content and expression. As in \system{}, we create synthetic review documents to cover item attributes. We do not compare methods beyond our scope, such as conversation strategy techniques based on predefined attributes or models that require user behavioral history or in-domain training data.

\paragraph{Implementation Details}
We used GPT-4o-mini for snippet extraction, and experimented with two LLMs as the \textit{recommender} system: OpenAI's GPT-4o-mini and LLaMa-3.3 (70B parameters). We tuned the hyperparameters and prompts of \system{} and baselines based on Hits@10 scores from the validation set. We tuned the number of retrieved snippets $k \in \{100, 500, 1000\}$ on validation set to find the best performing $k$ for different experiment setting. We used BGE~\cite{xiao2024c-pack}\footnote{\url{https://huggingface.co/BAAI/bge-base-en-v1.5}} for dense retrieval and NLI model by ~\citet{nie-etal-2020-adversarial} with $t_\text{entailment}=0.2$ for post-checking. These components were not fine-tuned during the experiments and are interchangeable with similar models. Experiments were run on a machine with 8 NVIDIA A100 GPUs, mainly for the NLI model and vector search engine. Each turn took around 5–10 seconds with a single GPU, using approximately 10GB of GPU memory. Appendix \ref{apx:infer-time} provides a detailed breakdown of runtime.

\paragraph{Evaluation Protocol} 
{We conduct simulated conversations up to five turns. To ensure that the user has an initial prompt and that the first step effectively narrows down the scope of reranking, the recommender starts with a fixed question: \texttt{``Hello, what category of restaurant are you looking for?''}, or \texttt{``Hello, what category of books/clothing items are you looking for?''}. The seeker is limited to responding to the provided questions. Based on the seeker's response, the recommender evaluates candidate items and calculates ranking metrics. At each subsequent turn, the recommender asks a brief clarification question. We compute Hits@k ($k = 1, 5, \text{and } 10$) and Mean Reciprocal Rank (MRR) at each turn. In cases of tied ranking, a random rank is assigned within the tied group, following~\citet{sun2020re}.}

\begin{table*}[t]
\centering
\small
\begin{tabular}{lrl|c@{\ \ }c|c@{\ \ }c}
\toprule
  & & & \multicolumn{2}{c}{\textbf{GPT-4o-mini}} & \multicolumn{2}{c}{\textbf{LLaMA-3.3 70B}} \\
        \midrule
       \textbf{Dataset} &
       \multicolumn{2}{c}{\textbf{Method}} & \textbf{Hits@10} & \textbf{MRR} & \textbf{Hits@10} & \textbf{MRR} \\
        \midrule
        \multirow{5}{*}{Yelp}
        & Baseline & Document & $0.283_{\ [0.26-0.31]}$ & $0.148_{\ [0.13-0.17]}$ & $0.257_{\ [0.23-0.28]}$ & $0.131_{\ [0.11-0.15 ]}$ \\
        &  & Sentence & $0.386_{\ [0.36-0.42]}$ & $0.231_{\ [0.21-0.25]}$ & $0.396_{\ [0.37-0.43]}$ & $0.227_{\ [0.21-0.25 ]}$ \\
        &  & Snippet & $0.414_{\ [0.38-0.44]}$ & $0.257_{\ [0.23-0.28]}$ & $0.415_{\ [0.38-0.45]}$ & $0.255_{\ [0.23-0.28 ]}$ \\
        & \system & Snippet & $0.454_{\ [0.42-0.48]}$ & $\textbf{0.277}_{\ [0.25-0.30]}$ & $0.438_{\ [0.41-0.47]}$ & $0.263_{\ [0.25-0.30 ]}$ \\
        &  & + Expansion (\S\ref{sec:snippet_user}) & $\textbf{0.464}_{\ [0.43-0.49]}$ & $\textbf{0.277}_{\ [0.25-0.30]}$ & $\textbf{0.459}_{\ [0.43-0.49]}$ & $\textbf{0.278}_{\ [0.24-0.29]}$ \\
        \midrule
        \multirow{5}{*}{Books}
        & Baseline & Document & $0.422_{\ [0.39-0.45]}$ & $0.227_{\ [0.21-0.25]}$ & $0.360_{\ [0.33-0.39]}$ & $0.209_{\ [0.19-0.23]}$ \\
        &  & Sentence & $0.479_{\ [0.45-0.51]}$ & $0.364_{\ [0.34-0.39]}$ & $0.422_{\ [0.39-0.45]}$ & $0.325_{\ [0.30-0.35 ]}$ \\
        &  & Snippet & $0.504_{\ [0.47-0.54]}$ & $0.385_{\ [0.36-0.41]}$ & $0.453_{\ [0.42-0.48]}$ & $0.349_{\ [0.32-0.38 ]}$ \\
        & \system{} & Snippet & $\textbf{0.552}_{\ [0.52-0.58]}$ & $\textbf{0.429}_{\ [0.39-0.44]}$ & $0.495_{\ [0.46-0.53]}$ & $0.376_{\ [0.40-0.46]}$ \\
        &  & + Expansion (\S\ref{sec:snippet_user}) & $0.545_{\ [0.51-0.58]}$ & $0.414_{\ [0.36-0.42]}$ & $\textbf{0.513}_{\ [0.48-0.54]}$ & $\textbf{0.389}_{\ [0.35-0.40]}$ \\
        \midrule
        \multirow{5}{*}{Clothing}
        & Baseline & Document & $0.167_{\ [0.14-0.19]}$ & $0.098_{\ [0.08-0.11]}$ & $0.170_{\ [0.15-0.19]}$ & $0.099_{\ [0.08-0.11]}$ \\
        &  & Sentence & $0.192_{\ [0.17-0.22]}$ & $0.125_{\ [0.11-0.14]}$ & $0.191_{\ [0.17-0.22]}$ & $0.123_{\ [0.10-0.14 ]}$ \\
        &  & Snippet & $0.212_{\ [0.19-0.24]}$ & $0.142_{\ [0.12-0.16]}$ & $0.208_{\ [0.18-0.23]}$ & $0.141_{\ [0.12-0.16 ]}$ \\
        & \system{} & Snippet & $\textbf{0.253}_{\ [0.23-0.28]}$ & $0.146_{\ [0.13-0.16]}$ & $0.249_{\ [0.22-0.28]}$ & $0.143_{\ [0.12-0.16]}$ \\
        &  & + Expansion (\S\ref{sec:snippet_user}) & $0.232_{\ [0.21-0.26]}$ & $\textbf{0.152}_{\ [0.13-0.17]}$ & $\textbf{0.254}_{\ [0.23-0.28]}$ & $\textbf{0.154}_{\ [0.14-0.17 ]}$ \\
        \bottomrule
    \end{tabular}
\caption{\textbf{Main results.} Hits@10 and MRR after five turns are reported with 95\% confidence intervals in brackets.}
\label{tb:final}
\end{table*}

\subsection{Results}

\paragraph{Main Results:}

{Table~\ref{tb:final} presents the main results, demonstrating the effectiveness of snippet-based representations. We observe that more fine-grained representations (from document-level to sentence-level, and to snippet-level) lead to improved retrieval performance in the baseline systems for both GPT and LLaMA. For example, the Hits@10 scores improved from 0.283 to 0.414 on the restaurant dataset, from 0.422 to 0.504 on the book dataset, and from 0.167 to 0.212 on the clothing dataset for GPT. \system{} outperformed the baseline by a clear margin with both models due to the advanced processing of query snippets~(\S\ref{sec:snippet_user}).

\paragraph{Analysis:} \naoki{While query expansion (\S\ref{sec:snippet_user}) improved performance consistently for LLaMA-based systems, it had mixed effects on GPT-based systems. Based on comparison of some examples, we conjecture that this may stem from differences in content diversity and structure. GPT tends to generate more formulaic item attributes (e.g., expanding ``The book is a children's book'' to ``The book features engaging activities for young readers''). LLaMA, though structurally similar, creates more unique and concrete descriptions (e.g., ``The story features talking animals.''). The ideal level of creativity may differ across domains, as indicated by the restaurant domain results. This suggests future research opportunities to optimize snippet expansion.}

\begin{figure}
    \centering
    \includegraphics[width=\linewidth]{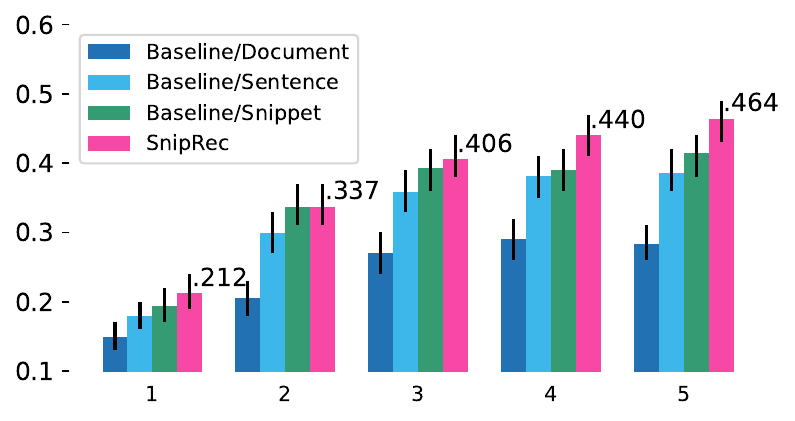}
    \caption{Hits@10 (vertical axis) over five turns (horizontal axis) in the restaurant dataset (GPT-4o-mini). The numbers represent \system{} scores. Similar trends were observed in other datasets for both GPT and LLaMA. Appendix~\ref{sec:appendix_full_comparison} presents full results.}
    \label{fig:turn_by_turn_scores}
\end{figure}

\paragraph{Progression Over Turns:}
\naoki{We observed that item retrieval performance clearly improved over five conversation turns on the restaurant dataset Figure~\ref{fig:turn_by_turn_scores}. Hits@10 for \system{} increased by 0.252 from the first turn to the fifth. Similar trends were noted in other datasets with both GPT and LLaMA.\footnote{Full results are available in Appendix~\ref{sec:appendix_full_comparison}. Conversation examples are provided in Appendix \ref{sec:interaction}.} This finding corroborates recent studies reporting that LLMs can function as robust conversation modules.}

\subsection{Quality of Extracted Snippets}\label{sec:snippet_evaluation}

We assessed a subset of extracted item snippets using both automatic and manual analyses, focusing on the faithfulness and atomicity of the snippets. Following \citet{wanner2024closer}, we employed the LLM-as-judge approach~\citet{zheng2023judging} to verify faithfulness--whether the original review supports the extracted snippets. This approach was applied to 1,000 snippet samples using GPT-4o-mini. We then manually reviewed a smaller set (100 snippets) to evaluate their atomicity alongside faithfulness. Two annotators reviewed the samples and defined two aspects of atomicity: a) whether each snippet addresses only one aspect of the item and b) whether each snippet discusses only one indivisible proposition. The annotators independently evaluated the same 100 samples and resolved disagreements through discussion.\footnote{Appendix~\ref{item-snippet-hallucination} provides further details.}

\begin{table}[t]
    \centering
    \small
    \begin{tabular}{@{\ }l@{\ \ }p{0.42\textwidth}}
    \multicolumn{2}{l}{\footnotesize \textbf{Atomic and Faithful Snippet} (94/100)} \\
    \toprule
    $R$ & [...] a great cookbook [...] you'll actually learn more about why recipes are constructed like they are. \\
    $s$ & The book provides explanations for recipe components. \\
    \midrule
    \multicolumn{2}{l}{\footnotesize \textbf{Hallucination} (3/100)} \\
    \toprule
    $R$ & [...] I ended up buying their last four bottles, so they hooked me up with a 10\% discount. \\
    $s$ & Discount is offered on bulk purchases. \\
    \midrule
    \multicolumn{2}{l}{\footnotesize \textbf{Atomic Snippets at the Aspect-level} (97/100)} \\
    \toprule
    $R$ & [...] Yeah the area is sketchy but their lot is kind of fenced in and the folks there seem fine [...] \\
    $s$ & The area is sketchy, but the lot is fenced in and feels safe. \\
    \bottomrule
    \end{tabular}
    \caption{Examples of evaluated snippets. $R$ and $s$ denote a review and a snippet, respectively.}
    \label{tab:snippet_evaluation}
\end{table}

We found that hallucinations are rare, with only about 3-3\% of item snippets judged unfaithful in both automatic and manual evaluations. Manual inspection showed that 97/100 item snippets are atomic regarding the aspects they describe, though 43/100 contain multiple propositions. Table~\ref{tab:snippet_evaluation} includes some annotated examples. Nevertheless, our main experimental results suggest that the observed level of atomicity is effective for item retrieval.

\subsection{Reliability of User Simulator}\label{sec:user-simulator-evaluation}

To assess the reliability of our user simulator, five annotators manually evaluated 50 sample responses across three dimensions: (1) relevance, (2) topicality, and (3) faithfulness to context. These dimensions were rated using binary scores, and judgments were aggregated by majority vote. The detailed procedure is described in Appendix~\ref{sec:appendix_simulator_humaneval}.

The number of examples judged as relevant, on-topic, and faithful were 50/50, 48/50, and 46/50, respectively, indicating the reliability of the user simulator. We observed that while GPT-4o-mini, the LLM model used in our simulator, generally follows instructions to provide relevant and on-topic answers, it may sometimes produce responses not fully grounded in the provided item information, particularly when questions or reviews contain rhetorical elements. We show examples in Appendix~\ref{sec:appendix_simulator}.

\section{Related Work}\label{sec:related-work}


\paragraph{CRS}
Many studies focus on data-driven conversation strategies~\cite{jannach2021survey}. These methods effectively elicit user preferences but are often limited by fixed predefined attributes~(e.g., \citealp{zhang2018towards,lei2020estimation}) or rely on expensive conversation data~(e.g., \citealp{li2018towards,hayati2020inspired}). In contrast, our work addresses a different challenge by representing item information and user preferences through snippets extracted from UGC. This approach can capture a broader range of user expressions across various domains.

\paragraph{UGC}
\naoki{Previous work has incorporated customer reviews in recommender systems through fine-tuned black-box embeddings, which often obscure the interpretability of the extracted information~\cite{sachdeva2019how,lu2021revcore}. Recent work incorporates review data directly into LLM-based systems to produce (non-conversational) recommendations~\cite{xu2024personalized,lyu2024llm}. Inspired by advances outside recommendation tasks \cite{min2023factscore, wanner2024closer}, we extract explicit snippets from UGC to improve item retrieval in CRS.}

paragraph{LLMs in CRS} 
\naoki{Recent work demonstrates the strength of LLMs as zero-shot agents for managing dialogues in CRS~\cite{he2023large, wang2023zero,friedman2023leveraging,wang2023rethinking}. Other studies highlight the strong question-asking capabilities of LLMs in different scenarios \cite{li2023eliciting,zhang2024probing}. LLMs have also enabled a new paradigm for simulating realistic conversations at scale~\cite{wang2023rethinking,yoon2024evaluating,liang2024llm,kim2024pearl}. These findings motivate us to delegate both the conversation strategy and user simulation to LLMs, allowing our work to focus on representation through few-shot snippet extraction.}

\section{Conclusion}\label{sec:conclusion}

In this paper, we propose leveraging UGC to capture a wide range of user preferences and effectively address the challenges faced by systems based on predefined attributes. \system{} utilizes snippets mined from customer reviews along with the implicit knowledge of LLMs to to accurately identify relevant items for free-form user queries. Our approach is resource-efficient and adaptable to various domains, as demonstrated by our experiments. The results also confirm the reliability of LLM-based snippet extraction and the user simulator for evaluation.


This work opens several avenues for future research. For instance, as our analysis suggests, the processes of snippet extraction, expansion, and retrieval could be refined to better model user preferences (e.g., by considering the certainty associated with each snippet and how snippet expansion and retrieval can be adapted to conversational contexts, or what kinds of snippet expansion contribute better to conversational recommendation performance). Given the reliability of the user simulator demonstrated in this study, these modules could also be fine-tuned using user simulation as the reward model. Furthermore, our problem setting could be extended to capture more complex user behaviors. Currently, our user simulator assumes that users have a single target item in mind. Potential alternative setting could involve scenarios where users are interested in multiple target items that share some characteristics, or that the user has more diverse intents other than seeking for particular items.

\section*{Limitations}

\paragraph{Interaction Type:} Some interaction types, such as providing feedback or explanations about recommendations, fall beyond our problem setting, where the recommender solely asks questions. This form of interaction is common in the early stages of human-to-human dialogue~\cite{lyu2021workflow}, but exploring the applicability of snippet-based representations to other types, such as social explanations~\cite{pecune2019model}, is an interesting direction for future work.

\paragraph{Reliance on UGC:} The proposed method relies on the availability and quality of user-generated content. Many domains such as restaurants, e-commerce, and movies have abundant UGC. However, in domains or languages where UGC is scarce or of low quality, the applicability and performance of \system{} may be limited.

\paragraph{Reliance on LLMs:} Our approach uses LLMs, which have limitations including high computational cost, response latency, hallucination, and potential biases. We provide empirical results to quantify these issues and demonstrate the method's utility~(\S\ref{sec:experiments}). Although recent studies highlight the utility of LLMs as proxies for human users (\S\ref{sec:related-work}), future research should explore how well they align with real users in depth. Addressing these concerns presents a potential avenue for future research.




\section*{Ethical Considerations}

Several ethical factors should be considered when using the proposed system, particularly regarding the generation of clarification questions. Since the system relies on LLMs, its output may reflect biases in the training data. For example, in response to a general query like ``I'm looking for Asian food,'' the system might suggest stereotypical dishes such as dumplings or sushi due to skewed data distributions. Additionally, there is a risk of the system generating harmful questions. While safeguard techniques can mitigate these issues to some extent~\cite{wang2024self}, caution is still required.

\bibliography{custom}

\appendix
\lstset{
  basicstyle=\ttfamily,
  frame=single,
  breaklines=true,
  showstringspaces=false,
  xleftmargin=2em,
  xrightmargin=2em,
  aboveskip=1em,
  belowskip=1em
}


\newtcolorbox{promptbox}[1]{width=\linewidth,
boxsep=2pt,left=8pt,right=7pt,top=5pt,bottom=5pt,
fontupper=\ttfamily,fontlower=\ttfamily,
fonttitle=\hypersetup{linkcolor=white,urlcolor=white},
title={#1},
}


\section{Additional Details}
\label{sec:appendix_detail}

\subsection{Item Snippets from Pre-defined attributes}
\label{sec:attribute_snippet}
In addition to extracting snippets from item reviews, we also generate item snippets based on pre-defined attributes, such as item categories or domain-specific features, using structured templates.
For categories, we apply the template ``\textsf{This place is categorized as \{category\}.}''
For other attributes, we use the format ``\textsf{it has \{attribute\} as \{value\}.}''
An example snippet would be 
\snippet{it has price range as under \$10.}




\subsection{Human Evaluation of User Simulator}
\label{sec:appendix_simulator_humaneval}
We collected 10 conversations, each consisting of 5 turns, resulting in a total of 50 simulator-generated responses using one of our baseline recommender systems. Five authors, all of whom work regularly with LLMs, have experience in NLP research, and are fluent in English, participated in annotating the responses.

\paragraph{Guidelines:} The annotators were provided with clear instructions, which included the definition of an effective user simulator: it should accurately answer the asked question without adding unnecessary information and remain consistent with the provided context, such as the review or item details. Additionally, the guidelines contains metric definitions (see below) and annotated examples for clarity. The annotation criteria were developed through annotation practice and discussions among the annotators.

\begin{itemize}
    \item Relevant: Does user's answer address system's question fully and appropriately, without omitting information? For instance, if system asks about food but user neither mentions a relevant preference at all nor explicitly states there is no preference, the response is not relevant.
    \item On-Topic: Does user's answer stay limited to the topic asked by system, without providing extra information? For instance, if system asks about food and user provides information other than food, it’s off-topic. If answered topics covered more than asked topics, it is likely off-topic.
    \item Faithful: Is user's answer supported by the provided information? Stating ``no preference'' even if the information contains preferences is considered inaccurate.
\end{itemize}

\paragraph{Inter-Annotator Agreement:} The distribution of collected judgments was highly skewed, making the use of widely used metrics like Fleiss' Kappa inappropriate for quantifying inter-annotator agreements. Therefore, we used Gwet's AC1, a method known for its robustness to class imbalance. The scores were 98.37 (Relevant), 97.48 (On-Topic), and 84.55 (Faithful), showing strong agreements among the annotators.


\section{Additional Experiment Results}
\label{sec:appendix_result}
In this section, we include further experimental results that are not included in the main manuscript due to space restrictions.

\subsection{Full Performance Comparison}
\label{sec:appendix_full_comparison}

Tables~\ref{tb:fullturnyelp}, \ref{tb:fullturnamazon} and \ref{tb:fullturnamazon-clothing} provide a comprehensive overview of the system performance across all compared methods.

\begin{table*}[htbp]
\tiny
\centering
\begin{tabular}{c|cccccccccc}
\toprule
\textbf{Turn} & \textbf{Hits@1} & \textbf{Hits@5} & \textbf{Hits@10} & \textbf{Avg Pos} & \textbf{Avg MRR} & \textbf{Hits@1 95ci} & \textbf{Hits@5 95ci} & \textbf{Hits@10 95ci} & \textbf{Avg Pos 95ci} & \textbf{Avg MRR 95ci} \\
\midrule
\multicolumn{11}{c}{\textbf{Baseline - Document} GPT-4o-mini (k=100)} \\
\midrule
1 & 0.020 & 0.091 & 0.149 & 757.709 & 0.065 & 0.01-0.03 & 0.07-0.11 & 0.13-0.17 & 694.98-820.44 & 0.06-0.08 \\
2 & 0.054 & 0.127 & 0.206 & 511.415 & 0.105 & 0.04-0.07 & 0.11-0.15 & 0.18-0.23 & 457.44-565.39 & 0.09-0.12 \\
3 & 0.077 & 0.169 & 0.270 & 364.596 & 0.135 & 0.06-0.09 & 0.15-0.19 & 0.24-0.30 & 319.54-409.65 & 0.12-0.15 \\
4 & 0.084 & 0.170 & 0.290 & 337.750 & 0.144 & 0.07-0.10 & 0.15-0.19 & 0.26-0.32 & 293.83-381.67 & 0.13-0.16 \\
5 & 0.089 & 0.167 & 0.283 & 307.381 & 0.148 & 0.07-0.11 & 0.14-0.19 & 0.26-0.31 & 266.76-348.01 & 0.13-0.17 \\
\midrule
\multicolumn{11}{c}{\textbf{Baseline - Document} LLaMA-3.3 70B (k=100)} \\
\midrule
1 & 0.018 & 0.090 & 0.153 & 704.959 & 0.065 & 0.01-0.03 & 0.07-0.11 & 0.13-0.18 & 646.76-763.16 & 0.06-0.08 \\
2 & 0.041 & 0.093 & 0.167 & 575.275 & 0.081 & 0.03-0.05 & 0.07-0.11 & 0.14-0.19 & 520.48-630.07 & 0.07-0.09 \\
3 & 0.055 & 0.146 & 0.230 & 418.054 & 0.110 & 0.04-0.07 & 0.12-0.17 & 0.20-0.26 & 371.06-465.05 & 0.10-0.12 \\
4 & 0.067 & 0.153 & 0.246 & 362.794 & 0.123 & 0.05-0.08 & 0.13-0.18 & 0.22-0.27 & 317.75-407.84 & 0.11-0.14 \\
5 & 0.074 & 0.167 & 0.257 & 337.193 & 0.131 & 0.06-0.09 & 0.14-0.19 & 0.23-0.28 & 294.50-379.89 & 0.11-0.15 \\
\midrule
\multicolumn{11}{c}{\textbf{Baseline - Sentence} GPT-4o-mini (k=500)} \\
\midrule
1 & 0.029 & 0.110 & 0.180 & 159.568 & 0.082 & 0.02-0.04 & 0.09-0.13 & 0.16-0.20 & 137.92-181.22 & 0.07-0.09 \\
2 & 0.083 & 0.220 & 0.299 & 116.661 & 0.158 & 0.07-0.10 & 0.19-0.25 & 0.27-0.33 & 98.23-135.09 & 0.14-0.18 \\
3 & 0.126 & 0.255 & 0.359 & 104.103 & 0.202 & 0.11-0.15 & 0.23-0.28 & 0.33-0.39 & 86.23-121.97 & 0.18-0.22 \\
4 & 0.149 & 0.285 & 0.381 & 89.070 & 0.222 & 0.13-0.17 & 0.26-0.31 & 0.35-0.41 & 73.23-104.91 & 0.20-0.24 \\
5 & 0.161 & 0.280 & 0.386 & 78.523 & 0.231 & 0.14-0.18 & 0.25-0.31 & 0.36-0.42 & 66.30-90.75 & 0.21-0.25 \\
\midrule
\multicolumn{11}{c}{\textbf{Baseline - Sentence} LLaMA-3.3 70B (k=100)} \\
\midrule
1 & 0.029 & 0.110 & 0.181 & 236.378 & 0.083 & 0.02-0.04 & 0.09-0.13 & 0.16-0.20 & 203.85-268.91 & 0.07-0.09 \\
2 & 0.078 & 0.170 & 0.238 & 197.668 & 0.136 & 0.06-0.09 & 0.15-0.19 & 0.21-0.26 & 168.19-227.15 & 0.12-0.15 \\
3 & 0.112 & 0.227 & 0.332 & 145.162 & 0.180 & 0.09-0.13 & 0.20-0.25 & 0.30-0.36 & 122.62-167.70 & 0.16-0.20 \\
4 & 0.138 & 0.273 & 0.380 & 118.890 & 0.215 & 0.12-0.16 & 0.25-0.30 & 0.35-0.41 & 99.23-138.55 & 0.19-0.24 \\
5 & 0.150 & 0.288 & 0.396 & 98.083 & 0.227 & 0.13-0.17 & 0.26-0.32 & 0.37-0.43 & 82.44-113.72 & 0.21-0.25 \\
\midrule
\multicolumn{11}{c}{\textbf{Baseline - Snippet} GPT-4o-mini (k=1000)} \\
\midrule
1 & 0.043 & 0.126 & 0.194 & 134.714 & 0.097 & 0.03-0.06 & 0.11-0.15 & 0.17-0.22 & 118.15-151.27 & 0.08-0.11 \\
2 & 0.109 & 0.231 & 0.337 & 95.198 & 0.181 & 0.09-0.13 & 0.20-0.26 & 0.31-0.37 & 80.84-109.56 & 0.16-0.20 \\
3 & 0.148 & 0.286 & 0.393 & 90.195 & 0.226 & 0.13-0.17 & 0.26-0.31 & 0.36-0.42 & 74.67-105.72 & 0.20-0.25 \\
4 & 0.171 & 0.294 & 0.390 & 78.418 & 0.243 & 0.15-0.19 & 0.27-0.32 & 0.36-0.42 & 66.24-90.60 & 0.22-0.27 \\
5 & 0.190 & 0.305 & 0.414 & 74.093 & 0.257 & 0.17-0.21 & 0.28-0.33 & 0.38-0.44 & 62.37-85.82 & 0.23-0.28 \\
\midrule
\multicolumn{11}{c}{\textbf{Baseline - Snippet} LLaMA-3.3 70B (k=100)} \\
\midrule
1 & 0.034 & 0.122 & 0.191 & 204.307 & 0.090 & 0.02-0.05 & 0.10-0.14 & 0.17-0.22 & 176.03-232.58 & 0.08-0.10 \\
2 & 0.085 & 0.191 & 0.271 & 174.967 & 0.148 & 0.07-0.10 & 0.17-0.22 & 0.24-0.30 & 148.24-201.69 & 0.13-0.17 \\
3 & 0.118 & 0.263 & 0.362 & 131.685 & 0.195 & 0.10-0.14 & 0.24-0.29 & 0.33-0.39 & 111.34-152.03 & 0.17-0.21 \\
4 & 0.160 & 0.290 & 0.400 & 109.844 & 0.233 & 0.14-0.18 & 0.26-0.32 & 0.37-0.43 & 91.10-128.59 & 0.21-0.25 \\
5 & 0.185 & 0.316 & 0.415 & 96.587 & 0.255 & 0.16-0.21 & 0.29-0.34 & 0.38-0.45 & 80.43-112.74 & 0.23-0.28 \\
\midrule
\multicolumn{11}{c}{\textbf{\system{}} GPT-4o-mini (k=500)} \\
\midrule
1 & 0.034 & 0.119 & 0.191 & 152.083 & 0.088 & 0.02-0.05 & 0.10-0.14 & 0.17-0.22 & 133.05-171.12 & 0.08-0.10 \\
2 & 0.097 & 0.218 & 0.305 & 111.660 & 0.166 & 0.08-0.12 & 0.19-0.24 & 0.28-0.33 & 94.83-128.49 & 0.15-0.18 \\
3 & 0.152 & 0.288 & 0.387 & 93.688 & 0.228 & 0.13-0.17 & 0.26-0.32 & 0.36-0.42 & 79.16-108.22 & 0.21-0.25 \\
4 & 0.180 & 0.315 & 0.429 & 81.767 & 0.257 & 0.16-0.20 & 0.29-0.34 & 0.40-0.46 & 68.26-95.27 & 0.23-0.28 \\
5 & 0.197 & 0.341 & 0.454 & 77.189 & 0.277 & 0.17-0.22 & 0.31-0.37 & 0.42-0.48 & 63.21-91.17 & 0.25-0.30 \\
\midrule
\multicolumn{11}{c}{\textbf{\system{}} LLaMA-3.3 70B (k=500)} \\
\midrule
1 & 0.032 & 0.127 & 0.209 & 143.613 & 0.091 & 0.02-0.04 & 0.11-0.15 & 0.18-0.23 & 126.61-160.61 & 0.08-0.10 \\
2 & 0.089 & 0.201 & 0.282 & 154.001 & 0.156 & 0.07-0.11 & 0.18-0.23 & 0.25-0.31 & 134.28-173.72 & 0.14-0.17 \\
3 & 0.131 & 0.270 & 0.366 & 135.404 & 0.205 & 0.11-0.15 & 0.24-0.30 & 0.34-0.40 & 114.79-156.02 & 0.18-0.23 \\
4 & 0.160 & 0.326 & 0.424 & 104.493 & 0.244 & 0.14-0.18 & 0.30-0.36 & 0.39-0.45 & 87.90-121.08 & 0.22-0.27 \\
5 & 0.185 & 0.333 & 0.438 & 93.701 & 0.263 & 0.16-0.21 & 0.30-0.36 & 0.41-0.47 & 78.24-109.16 & 0.24-0.29 \\
\midrule
\multicolumn{11}{c}{\textbf{\system{} + Expansion} GPT-4o-mini (k=500)} \\
\midrule
1 & 0.037 & 0.132 & 0.212 & 129.337 & 0.096 & 0.03-0.05 & 0.11-0.15 & 0.19-0.24 & 113.89-144.79 & 0.08-0.11 \\
2 & 0.090 & 0.229 & 0.337 & 98.499 & 0.169 & 0.07-0.11 & 0.20-0.26 & 0.31-0.37 & 84.04-112.95 & 0.15-0.19 \\
3 & 0.143 & 0.306 & 0.406 & 92.005 & 0.227 & 0.12-0.16 & 0.28-0.33 & 0.38-0.44 & 76.01-108.00 & 0.21-0.25 \\
4 & 0.172 & 0.346 & 0.440 & 75.537 & 0.260 & 0.15-0.20 & 0.32-0.38 & 0.41-0.47 & 63.02-88.05 & 0.24-0.28 \\
5 & 0.189 & 0.360 & 0.464 & 65.323 & 0.277 & 0.16-0.21 & 0.33-0.39 & 0.43-0.49 & 54.61-76.03 & 0.25-0.30 \\
\midrule
\multicolumn{11}{c}{\textbf{\system{} + Expansion} LLaMA-3.3 70B (k=500)} \\
\midrule
1 & 0.038 & 0.138 & 0.223 & 139.549 & 0.100 & 0.03-0.05 & 0.12-0.16 & 0.20-0.25 & 122.82-156.28 & 0.09-0.11 \\
2 & 0.092 & 0.205 & 0.301 & 127.229 & 0.159 & 0.07-0.11 & 0.18-0.23 & 0.27-0.33 & 110.47-143.99 & 0.14-0.18 \\
3 & 0.137 & 0.280 & 0.381 & 116.490 & 0.214 & 0.12-0.16 & 0.25-0.31 & 0.35-0.41 & 98.69-134.29 & 0.19-0.24 \\
4 & 0.175 & 0.329 & 0.440 & 95.486 & 0.256 & 0.15-0.20 & 0.30-0.36 & 0.41-0.47 & 79.67-111.30 & 0.23-0.28 \\
5 & 0.199 & 0.341 & 0.459 & 83.463 & 0.278 & 0.17-0.22 & 0.31-0.37 & 0.43-0.49 & 69.99-96.94 & 0.25-0.30 \\
\bottomrule
\hline
\end{tabular}
\caption{Full system performance at each turn for the restaurant dataset.}
\label{tb:fullturnyelp}
\end{table*}

\begin{table*}[htbp]
\tiny
\centering
\begin{tabular}{c|cccccccccc}
\toprule
\textbf{Turn} & \textbf{Hits@1} & \textbf{Hits@5} & \textbf{Hits@10} & \textbf{Avg Pos} & \textbf{Avg MRR} & \textbf{Hits@1 95ci} & \textbf{Hits@5 95ci} & \textbf{Hits@10 95ci} & \textbf{Avg Pos 95ci} & \textbf{Avg MRR 95ci} \\
\midrule
\multicolumn{11}{c}{\textbf{Baseline - Document} GPT-4o-mini (k=100)} \\
\midrule
1 & 0.020 & 0.076 & 0.133 & 1979.563 & 0.063 & 0.01-0.03 & 0.06-0.09 & 0.11-0.15 & 1790.70-2168.43 & 0.05-0.07 \\
2 & 0.062 & 0.205 & 0.305 & 1201.196 & 0.138 & 0.05-0.08 & 0.18-0.23 & 0.28-0.33 & 1045.12-1357.27 & 0.12-0.15 \\
3 & 0.106 & 0.268 & 0.371 & 956.239 & 0.188 & 0.09-0.13 & 0.24-0.30 & 0.34-0.40 & 812.28-1100.20 & 0.17-0.21 \\
4 & 0.130 & 0.284 & 0.403 & 846.761 & 0.210 & 0.11-0.15 & 0.26-0.31 & 0.37-0.43 & 712.86-980.67 & 0.19-0.23 \\
5 & 0.148 & 0.292 & 0.422 & 792.695 & 0.227 & 0.13-0.17 & 0.26-0.32 & 0.39-0.45 & 658.59-926.80 & 0.21-0.25 \\
\midrule
\multicolumn{11}{c}{\textbf{Baseline - Document} LLaMA-3.3 70B (k=100)} \\
\midrule
1 & 0.016 & 0.080 & 0.147 & 1868.929 & 0.063 & 0.01-0.02 & 0.06-0.10 & 0.13-0.17 & 1686.99-2050.87 & 0.05-0.07 \\
2 & 0.062 & 0.174 & 0.295 & 1264.248 & 0.127 & 0.05-0.08 & 0.15-0.20 & 0.27-0.32 & 1103.34-1425.16 & 0.11-0.14 \\
3 & 0.100 & 0.230 & 0.321 & 1021.103 & 0.173 & 0.08-0.12 & 0.20-0.26 & 0.29-0.35 & 875.37-1166.84 & 0.15-0.19 \\
4 & 0.124 & 0.257 & 0.339 & 882.698 & 0.197 & 0.10-0.14 & 0.23-0.28 & 0.31-0.37 & 746.00-1019.40 & 0.18-0.22 \\
5 & 0.138 & 0.261 & 0.360 & 859.349 & 0.209 & 0.12-0.16 & 0.23-0.29 & 0.33-0.39 & 726.88-991.82 & 0.19-0.23 \\
\midrule
\multicolumn{11}{c}{\textbf{Baseline - Sentence} GPT-4o-mini (k=100)} \\
\midrule
1 & 0.103 & 0.225 & 0.299 & 1506.339 & 0.169 & 0.08-0.12 & 0.20-0.25 & 0.27-0.33 & 1333.48-1679.19 & 0.15-0.19 \\
2 & 0.209 & 0.351 & 0.415 & 1136.255 & 0.282 & 0.18-0.23 & 0.32-0.38 & 0.38-0.45 & 982.63-1289.88 & 0.26-0.31 \\
3 & 0.257 & 0.404 & 0.472 & 1021.802 & 0.327 & 0.23-0.28 & 0.37-0.43 & 0.44-0.50 & 869.68-1173.92 & 0.30-0.35 \\
4 & 0.285 & 0.413 & 0.486 & 946.894 & 0.349 & 0.26-0.31 & 0.38-0.44 & 0.45-0.52 & 803.36-1090.43 & 0.32-0.38 \\
5 & 0.306 & 0.421 & 0.479 & 825.236 & 0.364 & 0.28-0.33 & 0.39-0.45 & 0.45-0.51 & 692.67-957.80 & 0.34-0.39 \\
\midrule
\multicolumn{11}{c}{\textbf{Baseline - Sentence} LLaMA-3.3 70B (k=1000)} \\
\midrule
1 & 0.101 & 0.220 & 0.288 & 562.415 & 0.165 & 0.08-0.12 & 0.19-0.25 & 0.26-0.32 & 453.63-671.20 & 0.15-0.18 \\
2 & 0.185 & 0.335 & 0.408 & 360.629 & 0.259 & 0.16-0.21 & 0.31-0.36 & 0.38-0.44 & 282.28-438.98 & 0.24-0.28 \\
3 & 0.237 & 0.353 & 0.439 & 303.673 & 0.299 & 0.21-0.26 & 0.32-0.38 & 0.41-0.47 & 233.82-373.52 & 0.27-0.32 \\
4 & 0.263 & 0.352 & 0.429 & 289.565 & 0.317 & 0.24-0.29 & 0.32-0.38 & 0.40-0.46 & 222.26-356.87 & 0.29-0.34 \\
5 & 0.274 & 0.366 & 0.422 & 247.523 & 0.325 & 0.25-0.30 & 0.34-0.40 & 0.39-0.45 & 198.17-296.88 & 0.30-0.35 \\
\midrule
\multicolumn{11}{c}{\textbf{Baseline - Snippet} GPT-4o-mini (k=1000)} \\
\midrule
1 & 0.113 & 0.237 & 0.302 & 493.710 & 0.179 & 0.09-0.13 & 0.21-0.26 & 0.27-0.33 & 399.35-588.07 & 0.16-0.20 \\
2 & 0.222 & 0.352 & 0.428 & 350.655 & 0.291 & 0.20-0.25 & 0.32-0.38 & 0.40-0.46 & 274.10-427.21 & 0.27-0.32 \\
3 & 0.274 & 0.421 & 0.495 & 292.019 & 0.348 & 0.25-0.30 & 0.39-0.45 & 0.46-0.53 & 224.40-359.64 & 0.32-0.37 \\
4 & 0.305 & 0.429 & 0.499 & 320.447 & 0.369 & 0.28-0.33 & 0.40-0.46 & 0.47-0.53 & 246.70-394.19 & 0.34-0.40 \\
5 & 0.325 & 0.446 & 0.504 & 255.871 & 0.385 & 0.30-0.35 & 0.42-0.48 & 0.47-0.54 & 200.36-311.38 & 0.36-0.41 \\
\midrule
\multicolumn{11}{c}{\textbf{Baseline - Snippet} LLaMA-3.3 70B (k=500)} \\
\midrule
1 & 0.116 & 0.243 & 0.311 & 758.719 & 0.186 & 0.10-0.14 & 0.22-0.27 & 0.28-0.34 & 636.88-880.55 & 0.17-0.21 \\
2 & 0.205 & 0.361 & 0.431 & 568.457 & 0.282 & 0.18-0.23 & 0.33-0.39 & 0.40-0.46 & 462.04-674.87 & 0.26-0.31 \\
3 & 0.252 & 0.381 & 0.461 & 507.324 & 0.317 & 0.23-0.28 & 0.35-0.41 & 0.43-0.49 & 406.86-607.79 & 0.29-0.34 \\
4 & 0.276 & 0.384 & 0.451 & 452.367 & 0.334 & 0.25-0.30 & 0.35-0.41 & 0.42-0.48 & 357.41-547.32 & 0.31-0.36 \\
5 & 0.294 & 0.395 & 0.453 & 460.492 & 0.349 & 0.27-0.32 & 0.36-0.43 & 0.42-0.48 & 365.74-555.25 & 0.32-0.38 \\
\midrule
\multicolumn{11}{c}{\textbf{\system{}}  GPT-4o-mini (k=1000)} \\
\midrule
1 & 0.155 & 0.322 & 0.391 & 455.690 & 0.235 & 0.13-0.18 & 0.29-0.35 & 0.36-0.42 & 368.47-542.91 & 0.21-0.26 \\
2 & 0.259 & 0.430 & 0.492 & 328.777 & 0.339 & 0.23-0.29 & 0.40-0.46 & 0.46-0.52 & 255.68-401.87 & 0.31-0.36 \\
3 & 0.325 & 0.483 & 0.559 & 306.239 & 0.400 & 0.30-0.35 & 0.45-0.51 & 0.53-0.59 & 230.10-382.38 & 0.37-0.43 \\
4 & 0.357 & 0.485 & 0.550 & 251.113 & 0.420 & 0.33-0.39 & 0.45-0.52 & 0.52-0.58 & 190.80-311.42 & 0.39-0.45 \\
5 & 0.368 & 0.489 & 0.552 & 258.158 & 0.429 & 0.34-0.40 & 0.46-0.52 & 0.52-0.58 & 193.37-322.94 & 0.40-0.46 \\
\midrule
\multicolumn{11}{c}{\textbf{\system{}} LLaMA-3.3 70B (k=1000)} \\
\midrule
1 & 0.149 & 0.302 & 0.370 & 605.365 & 0.225 & 0.13-0.17 & 0.27-0.33 & 0.34-0.40 & 496.77-713.96 & 0.20-0.25 \\
2 & 0.240 & 0.389 & 0.466 & 401.158 & 0.314 & 0.21-0.27 & 0.36-0.42 & 0.44-0.50 & 316.25-486.06 & 0.29-0.34 \\
3 & 0.289 & 0.413 & 0.491 & 346.228 & 0.353 & 0.26-0.32 & 0.38-0.44 & 0.46-0.52 & 269.90-422.55 & 0.33-0.38 \\
4 & 0.314 & 0.420 & 0.491 & 326.659 & 0.369 & 0.29-0.34 & 0.39-0.45 & 0.46-0.52 & 254.03-399.29 & 0.34-0.40 \\
5 & 0.323 & 0.430 & 0.495 & 329.054 & 0.376 & 0.29-0.35 & 0.40-0.46 & 0.46-0.53 & 256.44-401.67 & 0.35-0.40 \\
\midrule
\multicolumn{11}{c}{\textbf{\system{} + Expansion} GPT-4o-mini (k=1000)} \\
\midrule
1 & 0.150 & 0.318 & 0.384 & 333.929 & 0.230 & 0.13-0.17 & 0.29-0.35 & 0.35-0.41 & 268.01-399.85 & 0.21-0.25 \\
2 & 0.246 & 0.414 & 0.489 & 276.166 & 0.329 & 0.22-0.27 & 0.38-0.44 & 0.46-0.52 & 214.54-337.79 & 0.30-0.35 \\
3 & 0.311 & 0.464 & 0.535 & 219.170 & 0.387 & 0.28-0.34 & 0.43-0.49 & 0.50-0.57 & 168.26-270.08 & 0.36-0.41 \\
4 & 0.335 & 0.471 & 0.540 & 225.405 & 0.401 & 0.31-0.36 & 0.44-0.50 & 0.51-0.57 & 172.79-278.02 & 0.37-0.43 \\
5 & 0.353 & 0.482 & 0.545 & 209.771 & 0.414 & 0.32-0.38 & 0.45-0.51 & 0.51-0.58 & 161.52-258.02 & 0.39-0.44 \\
\midrule
\multicolumn{11}{c}{\textbf{\system{} + Expansion} LLaMA-3.3 70B (k=1000)} \\
\midrule
1 & 0.152 & 0.321 & 0.394 & 340.011 & 0.234 & 0.13-0.17 & 0.29-0.35 & 0.36-0.42 & 271.78-408.24 & 0.21-0.26 \\
2 & 0.249 & 0.403 & 0.489 & 235.439 & 0.326 & 0.22-0.28 & 0.37-0.43 & 0.46-0.52 & 183.89-286.99 & 0.30-0.35 \\
3 & 0.291 & 0.438 & 0.520 & 208.621 & 0.362 & 0.26-0.32 & 0.41-0.47 & 0.49-0.55 & 162.37-254.87 & 0.34-0.39 \\
4 & 0.315 & 0.443 & 0.513 & 204.563 & 0.379 & 0.29-0.34 & 0.41-0.47 & 0.48-0.54 & 157.46-251.67 & 0.35-0.41 \\
5 & 0.331 & 0.439 & 0.513 & 197.468 & 0.389 & 0.30-0.36 & 0.41-0.47 & 0.48-0.54 & 158.00-236.94 & 0.36-0.42 \\
\bottomrule
\hline
\end{tabular}
\caption{Full system performance at each turn for the book dataset.}
\label{tb:fullturnamazon}
\end{table*}

\begin{table*}[htbp]
\tiny
\centering
\begin{tabular}{c|cccccccccc}
\toprule
\textbf{Turn} & \textbf{Hits@1} & \textbf{Hits@5} & \textbf{Hits@10} & \textbf{Avg Pos} & \textbf{Avg MRR} & \textbf{Hits@1 95ci} & \textbf{Hits@5 95ci} & \textbf{Hits@10 95ci} & \textbf{Avg Pos 95ci} & \textbf{Avg MRR 95ci} \\
\midrule
\multicolumn{11}{c}{\textbf{Baseline - Document} GPT-4o-mini (k=100)} \\
\midrule
1 & 0.010 & 0.050 & 0.107 & 1934.348 & 0.047 & 0.00-0.02 & 0.04-0.06 & 0.09-0.13 & 1750.77-2117.93 & 0.04-0.05 \\
2 & 0.035 & 0.092 & 0.133 & 1515.333 & 0.072 & 0.02-0.05 & 0.07-0.11 & 0.11-0.15 & 1352.60-1678.07 & 0.06-0.08 \\
3 & 0.042 & 0.100 & 0.150 & 1519.049 & 0.079 & 0.03-0.05 & 0.08-0.12 & 0.13-0.17 & 1354.95-1683.15 & 0.07-0.09 \\
4 & 0.052 & 0.110 & 0.163 & 1341.672 & 0.090 & 0.04-0.07 & 0.09-0.13 & 0.14-0.19 & 1179.50-1503.84 & 0.08-0.10 \\
5 & 0.059 & 0.118 & 0.167 & 1254.036 & 0.098 & 0.04-0.07 & 0.10-0.14 & 0.14-0.19 & 1099.20-1408.87 & 0.08-0.11 \\
\midrule
\multicolumn{11}{c}{\textbf{Baseline - Document} LLaMA-3.3 70B (k=100)} \\
\midrule
1 & 0.011 & 0.058 & 0.114 & 1875.500 & 0.050 & 0.00-0.02 & 0.04-0.07 & 0.09-0.13 & 1692.44-2058.56 & 0.04-0.06 \\
2 & 0.034 & 0.110 & 0.150 & 1574.288 & 0.076 & 0.02-0.05 & 0.09-0.13 & 0.13-0.17 & 1403.32-1745.26 & 0.06-0.09 \\
3 & 0.048 & 0.115 & 0.185 & 1351.646 & 0.091 & 0.03-0.06 & 0.10-0.13 & 0.16-0.21 & 1192.86-1510.43 & 0.08-0.10 \\
4 & 0.057 & 0.125 & 0.178 & 1234.909 & 0.099 & 0.04-0.07 & 0.10-0.15 & 0.15-0.20 & 1082.26-1387.56 & 0.08-0.11 \\
5 & 0.062 & 0.117 & 0.170 & 1242.217 & 0.099 & 0.05-0.08 & 0.10-0.14 & 0.15-0.19 & 1088.56-1395.87 & 0.08-0.11 \\
\midrule
\multicolumn{11}{c}{\textbf{Baseline - Sentence} GPT-4o-mini (k=1000)} \\
\midrule
1 & 0.031 & 0.113 & 0.170 & 780.693 & 0.082 & 0.02-0.04 & 0.09-0.13 & 0.15-0.19 & 653.85-907.53 & 0.07-0.09 \\
2 & 0.064 & 0.115 & 0.160 & 706.770 & 0.099 & 0.05-0.08 & 0.10-0.13 & 0.14-0.18 & 597.16-816.38 & 0.08-0.11 \\
3 & 0.078 & 0.132 & 0.186 & 653.231 & 0.114 & 0.06-0.09 & 0.11-0.15 & 0.16-0.21 & 553.00-753.46 & 0.10-0.13 \\
4 & 0.085 & 0.141 & 0.187 & 594.313 & 0.120 & 0.07-0.10 & 0.12-0.16 & 0.16-0.21 & 503.94-684.69 & 0.10-0.14 \\
5 & 0.090 & 0.138 & 0.192 & 564.795 & 0.125 & 0.07-0.11 & 0.12-0.16 & 0.17-0.22 & 481.93-647.66 & 0.11-0.14 \\
\midrule
\multicolumn{11}{c}{\textbf{Baseline - Sentence} LLaMA-3.3 70B (k=1000)} \\
\midrule
1 & 0.032 & 0.110 & 0.167 & 748.171 & 0.083 & 0.02-0.04 & 0.09-0.13 & 0.14-0.19 & 629.35-866.99 & 0.07-0.09 \\
2 & 0.053 & 0.112 & 0.158 & 704.295 & 0.094 & 0.04-0.07 & 0.09-0.13 & 0.14-0.18 & 596.25-812.34 & 0.08-0.11 \\
3 & 0.068 & 0.126 & 0.183 & 691.014 & 0.105 & 0.05-0.08 & 0.11-0.15 & 0.16-0.21 & 584.23-797.80 & 0.09-0.12 \\
4 & 0.083 & 0.133 & 0.191 & 631.332 & 0.117 & 0.07-0.10 & 0.11-0.15 & 0.17-0.22 & 536.46-726.21 & 0.10-0.13 \\
5 & 0.088 & 0.141 & 0.191 & 596.028 & 0.123 & 0.07-0.11 & 0.12-0.16 & 0.17-0.22 & 506.10-685.95 & 0.10-0.14 \\
\midrule
\multicolumn{11}{c}{\textbf{Baseline - Snippet} GPT-4o-mini (k=500)} \\
\midrule
1 & 0.039 & 0.128 & 0.200 & 1081.991 & 0.093 & 0.03-0.05 & 0.11-0.15 & 0.18-0.22 & 935.58-1228.40 & 0.08-0.11 \\
2 & 0.071 & 0.142 & 0.194 & 951.478 & 0.115 & 0.06-0.09 & 0.12-0.16 & 0.17-0.22 & 820.43-1082.52 & 0.10-0.13 \\
3 & 0.090 & 0.154 & 0.199 & 929.788 & 0.128 & 0.07-0.11 & 0.13-0.18 & 0.17-0.22 & 801.13-1058.45 & 0.11-0.15 \\
4 & 0.097 & 0.162 & 0.203 & 858.729 & 0.136 & 0.08-0.12 & 0.14-0.18 & 0.18-0.23 & 733.62-983.84 & 0.12-0.15 \\
5 & 0.107 & 0.162 & 0.212 & 771.016 & 0.142 & 0.09-0.13 & 0.14-0.18 & 0.19-0.24 & 656.00-886.03 & 0.12-0.16 \\
\midrule
\multicolumn{11}{c}{\textbf{Baseline - Snippet} LLaMA-3.3 70B (k=100)} \\
\midrule
1 & 0.045 & 0.133 & 0.207 & 2085.765 & 0.098 & 0.03-0.06 & 0.11-0.15 & 0.18-0.23 & 1895.93-2275.60 & 0.08-0.11 \\
2 & 0.079 & 0.139 & 0.185 & 1785.214 & 0.117 & 0.06-0.10 & 0.12-0.16 & 0.16-0.21 & 1610.68-1959.75 & 0.10-0.13 \\
3 & 0.093 & 0.152 & 0.193 & 1715.030 & 0.128 & 0.07-0.11 & 0.13-0.17 & 0.17-0.22 & 1540.30-1889.76 & 0.11-0.15 \\
4 & 0.103 & 0.152 & 0.205 & 1639.961 & 0.136 & 0.08-0.12 & 0.13-0.17 & 0.18-0.23 & 1466.08-1813.84 & 0.12-0.15 \\
5 & 0.108 & 0.156 & 0.208 & 1555.878 & 0.141 & 0.09-0.13 & 0.13-0.18 & 0.18-0.23 & 1382.13-1729.63 & 0.12-0.16 \\
\midrule
\multicolumn{11}{c}{\textbf{\system{}}  GPT-4o-mini (k=1000)} \\
\midrule
1 & 0.041 & 0.133 & 0.212 & 1048.920 & 0.099 & 0.03-0.05 & 0.11-0.15 & 0.19-0.24 & 906.72-1191.12 & 0.09-0.11 \\
2 & 0.062 & 0.156 & 0.227 & 840.274 & 0.121 & 0.05-0.08 & 0.13-0.18 & 0.20-0.25 & 709.96-970.59 & 0.10-0.14 \\
3 & 0.081 & 0.170 & 0.241 & 716.963 & 0.136 & 0.06-0.10 & 0.15-0.19 & 0.21-0.27 & 600.57-833.36 & 0.12-0.15 \\
4 & 0.091 & 0.168 & 0.232 & 634.158 & 0.142 & 0.07-0.11 & 0.14-0.19 & 0.21-0.26 & 527.93-740.39 & 0.12-0.16 \\
5 & 0.095 & 0.175 & 0.253 & 606.619 & 0.146 & 0.08-0.11 & 0.15-0.20 & 0.23-0.28 & 501.88-711.36 & 0.13-0.16 \\
\midrule
\multicolumn{11}{c}{\textbf{\system{}} LLaMA-3.3 70B (k=1000)} \\
\midrule
1 & 0.037 & 0.141 & 0.215 & 825.146 & 0.096 & 0.03-0.05 & 0.12-0.16 & 0.19-0.24 & 696.72-953.57 & 0.08-0.11 \\
2 & 0.069 & 0.165 & 0.233 & 708.483 & 0.124 & 0.05-0.08 & 0.14-0.19 & 0.21-0.26 & 589.36-827.61 & 0.11-0.14 \\
3 & 0.080 & 0.158 & 0.241 & 535.007 & 0.131 & 0.06-0.10 & 0.14-0.18 & 0.21-0.27 & 438.02-631.99 & 0.11-0.15 \\
4 & 0.083 & 0.164 & 0.250 & 507.373 & 0.135 & 0.07-0.10 & 0.14-0.19 & 0.22-0.28 & 415.12-599.63 & 0.12-0.15 \\
5 & 0.091 & 0.170 & 0.249 & 514.354 & 0.143 & 0.07-0.11 & 0.15-0.19 & 0.22-0.28 & 419.79-608.91 & 0.12-0.16 \\
\midrule
\multicolumn{11}{c}{\textbf{\system{} + Expansion} GPT-4o-mini (k=500)} \\
\midrule
1 & 0.040 & 0.128 & 0.208 & 979.550 & 0.097 & 0.03-0.05 & 0.11-0.15 & 0.18-0.23 & 841.81-1117.29 & 0.08-0.11 \\
2 & 0.065 & 0.162 & 0.231 & 759.479 & 0.123 & 0.05-0.08 & 0.14-0.18 & 0.20-0.26 & 640.86-878.10 & 0.11-0.14 \\
3 & 0.090 & 0.167 & 0.232 & 656.578 & 0.141 & 0.07-0.11 & 0.14-0.19 & 0.21-0.26 & 545.53-767.63 & 0.12-0.16 \\
4 & 0.101 & 0.168 & 0.226 & 576.390 & 0.149 & 0.08-0.12 & 0.14-0.19 & 0.20-0.25 & 475.24-677.54 & 0.13-0.17 \\
5 & 0.108 & 0.174 & 0.232 & 498.779 & 0.152 & 0.09-0.13 & 0.15-0.20 & 0.21-0.26 & 410.83-586.73 & 0.13-0.17 \\
\midrule
\multicolumn{11}{c}{\textbf{\system{} + Expansion} LLaMA-3.3 70B (k=1000)} \\
\midrule
1 & 0.033 & 0.136 & 0.218 & 558.727 & 0.094 & 0.02-0.04 & 0.11-0.16 & 0.19-0.24 & 464.37-653.09 & 0.08-0.11 \\
2 & 0.072 & 0.167 & 0.239 & 507.867 & 0.129 & 0.06-0.09 & 0.14-0.19 & 0.21-0.27 & 416.63-599.10 & 0.11-0.15 \\
3 & 0.086 & 0.173 & 0.246 & 439.198 & 0.140 & 0.07-0.10 & 0.15-0.20 & 0.22-0.27 & 359.31-519.08 & 0.12-0.16 \\
4 & 0.098 & 0.175 & 0.255 & 423.658 & 0.148 & 0.08-0.12 & 0.15-0.20 & 0.23-0.28 & 346.45-500.86 & 0.13-0.17 \\
5 & 0.109 & 0.177 & 0.254 & 426.539 & 0.154 & 0.09-0.13 & 0.15-0.20 & 0.23-0.28 & 349.63-503.45 & 0.14-0.17 \\
\bottomrule
\hline
\end{tabular}
\caption{Full system performance at each turn for the clothing dataset.}
\label{tb:fullturnamazon-clothing}
\end{table*}

\subsection{Data-aware Generation of Clarification Questions}
\label{sec:sampled}

As noted in the Limitations section, our study does not focus on advanced techniques for generating clarification questions, and \system{} relies on the internal knowledge of LLMs. To provide insights into directions for future research, we present an initial attempt to generate questions based on underlying data, motivated by recent work in conversational information retrieval~\cite{zhao2024generating}.




This approach seeks to identify distinctive facets within the current search space and generate clarifying questions based on these facets. As our knowledge source---a collection of snippets---is not fully structured or canonicalized, quantifying topics and computing their informativeness is not trivial. To address this, we cluster the snippets associated with the top candidates and dynamically identify representative topics within the current search space.

First, we focus on the top-$k_\text{item}=100$ candidates and collect their item snippets that have not yet been retrieved. From these, we sample 1,000 snippets. We then cluster these snippets into $C \in {25, 50, 100}$ clusters using a Gaussian Mixture Model (GMM).\footnote{We follow the technique employed in RAPTOR~\cite{sarthi2024raptor}, which involves dimensionality reduction using UMAP~\cite{mcinnes2018umap} before applying the GMM.} Inspired by the information gain score~\cite{zhao2024generating}, we rank the clusters based on the ratio of top-$k$ items linked to at least one snippet in the unexplored set $S^u$. Finally, we select the top-$k_\text{cluster}=10$ clusters to quantify the representative topics. For each cluster, we extract the three sentences closest to the cluster center. These groups of sentences form a summary of the underlying data. An example is shown below:
\begin{promptbox}{}
\scriptsize
- Topic group 1:\\
    - the place is a little cramped but fitting with the rest of the area.\\
    - seating outdoors was a little strange for two people, but nice for people watching.\\
    - the place is spotless.\\
- Topic group 2:\\
    - cakes and pies are delicious.\\
    - the desserts are loved wholeheartedly and are unforgettable.\\
    - desserts are fantastic, low carb or not.\\
- Topic group 3:\\
    - the bakery is worth visiting.\\
    - the bakery plans to return.\\
    - the bakery will still be a favorite in town, despite negative occurrences.\\
- Topic group 4:\\
    - service has improved significantly recently.\\
    - the service was above and beyond.\\
    - the delivery service is efficient and easy to arrange.\\
- Topic group 5:\\
    - employees were really friendly.\\
    - upon entering, i was immediately warmly greeted by a kind staff member at the door.\\
    - the staff is extremely helpful and greets customers immediately.\\
- Topic group 6:\\
    - chocolate cake is moist and rich.\\
    - butter cake is really enjoyed.\\
    - the rum cake is exceptional.\\
- Topic group 7:\\
    - the restaurant has over 800 reviews.\\
    - the restaurant is reliable.\\
    - overall, the restaurant has improved.\\
- Topic group 8:\\
    - the cake is reasonably priced.\\
    - the design of the wedding cake was perfect and just what was wanted.\\
    - mango cake has some glitter on top of it.\\
- Topic group 9:\\
    - strawberry nutella is a favored flavor.\\
    - homemade coconut syrup is used as a sweetener.\\
    - banana pudding is pretty good.\\
- Topic group 10:\\
    - the food quality is high.\\
    - the quality of food is good, deserving of four stars.\\
    - the food is fresh and good.
\end{promptbox}

We then use this summary as part of the instructions to generate clarification questions, as demonstrated in the following prompt:
\begin{promptbox}{Prompt for generating clarification questions based on candidate items}
\scriptsize
You are a Recommender chatting with a Seeker to provide restaurant/book recommendation. Your task is to ask questions for understanding the Seeker's preference.

\tcblower
\scriptsize
\# Role-Play Task: Recommender\\
You will play the role of a Recommender helping a Seeker find a restaurant/book that suits the Seeker's preferences.\\
\\
Based on the conversation log provided below, identify the most relevant aspect of the Seeker's preferences that will help refine the search for a suitable restaurant/book. Your question should focus only on one topic. Do not ask about multiple topics at once.\\
\\
**Topics Relevant to the Current Top Candidates:**\\
Consider the following topics associted with the current top candidates.\\
\{data\_summary\}\\
\\
**Dialogue Context:**\\
\{context\}\\
Recommender:\\
\\
Now, generate a response in the role of the Recommender.\\
\\
**Response Guidelines:**\\
\\
- Your response should be concise, typically one sentence. Avoid asking multiple questions at once.\\
- Do not ask for a restaurant name or any personal or street names. \\
Respond directly and concisely to the scenario without repeating the instructions or adding unrelated details. Use question types that give the human user more flexibility, allowing for creative and open-ended answers while staying relevant to the context.
\end{promptbox}

\begin{table}[th]
\centering
\small
\begin{tabular}{lccc}
\toprule
 & Hits@1 & Hits@10 & MRR \\
 \midrule
\system & 0.182 & 0.448 & 0.265 \\
+ Data Summary (\ref{sec:sampled}) & 0.153 & 0.343 & 0.214 \\
\bottomrule
\end{tabular}
\caption{Validation performance with and without the summary of data in input.}\label{tab:data-aware}
\end{table}

In our experiments on the validation set, this data-aware question generation method performed significantly worse than the LLM-based generation approach. Table~\ref{tab:data-aware} compares the performance of GPT-4o-mini-based recommenders with and without the data summaries. We observed that the data summaries often contained similar topics across different turns. These topics, such as service, price, and atmosphere, were frequently mentioned in many reviews across different items, but they do not necessarily convey key information needed to distinguish between similar items for recommendation purposes. Future research could explore more advanced techniques to identify key topics and better quantify their informativeness.

\subsection{Inference Time}
\label{apx:infer-time}

\system{} takes around 5–10 seconds per turn on a single GPU (see \S\ref{sec:exp-setting}). Table~\ref{tab:inference-time} provides a breakdown of the inference time. Over 70\% of the time is spent on the parser, expander, and responder components, mainly due to LLM API call latency. The validator accounts for roughly 13\% of the time, with its runtime varying based on the number of retrieved snippets and GPU performance. The retriever may slow down with more snippets, although modern dense retrievers like faiss are highly scalable.

\begin{table}
    \centering
    \small
    \begin{tabular}{lccc}
        \toprule
        \textbf{Module} & \textbf{Restaurants} & \textbf{Books} & \textbf{Clothing} \\
        \midrule
        parser\textsuperscript{API} & ${0.87}_{\ (\pm 0.55)}$ & ${1.08}_{\ (\pm 0.69)}$ & ${0.93}_{\ (\pm 0.52)}$ \\
        expander\textsuperscript{API} & ${2.28}_{\ (\pm 4.71)}$ & ${3.36}_{\ (\pm 2.50)}$ & ${3.29}_{\ (\pm 6.67)}$ \\
        retriever & ${0.59}_{\ (\pm 0.52)}$ & ${0.70}_{\ (\pm 0.70)}$ & ${1.01}_{\ (\pm 0.60)}$ \\
        validator\textsuperscript{GPU} & ${0.76}_{\ (\pm 0.53)}$ & ${0.99}_{\ (\pm 0.59)}$ & ${0.96}_{\ (\pm 0.59)}$ \\
        scorer & $<10_\text{ms}$ & $<10_\text{ms}$ & $<10_\text{ms}$ \\
        responder\textsuperscript{API} & ${0.68}_{\ (\pm 0.49)}$ & ${0.78}_{\ (\pm 0.57)}$ & ${0.65}_{\ (\pm 0.40)}$ \\
        \bottomrule
    \end{tabular}
    \caption{Breakdown of inference time (in seconds). The standard deviation is reported in parentheses. ``API'' and ``GPU'' refer to the modules whose runtime is primarily influenced by the latency of API calls and the performance of GPUs, respectively.}
    \label{tab:inference-time}
\end{table}

\section{List of Prompts}
\label{sec:prompts}

We provide the key prompts used in \system{}, with system prompts listed in the top section and user prompts in the bottom section. All prompts, along with the system implementation, are included in the supplementary material of this submission. Please refer to the README in the material for navigation.


\subsection{Item Snippet Extraction (\S~\ref{sec:snippet_review})}
\label{sec:prompt_item_snippet}

\begin{promptbox}{Prompt for decomposing reviews into item snippets (Yelp) }
\scriptsize
You are a helpful assistant. Follow the instructions. No prose.
\tcblower
\scriptsize
As a language genius, you are tasked with reading restaurant reviews and extracting and summarizing atomic, simple, short and coherent sentences that contain factual descriptions or subjective opinions related to a specific aspect of the restaurant, and subsequently you also need to (1) identify the topic of each atomic sentence and (2) the attitude towards the restaurant in terms of this proposition. The topics should be the aspects relevant to restaurant domains. You should use multiple atomic propositions if the content is about several topics, but combine similar content into one proposition.\\
\\
Ensure these sentences carry information that effectively help differentiate various restaurants. If there is already a proposition with a similar meaning, ignore the redundant information. Ignore the irrelevant chatter, narratives and descriptions unrelated to the properties of the restaurant in the reviews. Try to use original texts from the reviews and but do summarize them if they are verbose. Be sure to cover the whole review. Try to eliminate any references of "I" or "reviewer", but focus on "restaurant". Follow the exemplar format to extract.\\
\\
Analyze this review:\\
\`{}\`{}\`{}Harbor Seafood is the type of restaurant that's most like a bag of Lay's chips. One visit is never enough.  I can't begin to count the number of patrons I hear say they have just landed at the airport and headed straight over, or they have been in town a few days and already visited multiple times, or are headed home and just had to stop in one more time before the flight home.  The one thing I love most about this restaurant even more than the food is the fact that I can sit at the bar on any given day and I'm just as likely to sit next to a regular as I am to sit next to someone from half way around the world. People literally flock to this hole in the wall neighborhood seafood restaurant.  It's a relatively small restaurant so if you need a table your more than likely going to have to wait. Sometimes when it's real busy there is even a wait for the bar. All seating is first come first served \& they do not take reservations. You can also order food to go if you don't have the inclination or the time to sit in the restaurant and eat, but I'll admit this is the type of food that is at its best when you sit in and eat.\`{}\`{}\`{} \\
\\
{[\\
    "One visit is never enough."\\
    "Patrons love to visit this restaurant multiple times.",\\
    "Some guests come from halfway around the world.",\\
    "It's a relatively small restaurant.",\\
    "Waiting for a table is probable.",\\
    "Wait for the bar is also possible.",\\
    "All seating is first come, first served.",\\
    "They do not take reservations.",\\
    "Order food to go is allowed.",\\
    "Sit in and eat is recommended for this kind of food."\\
]}\\
\\
Analyze this review:\\
\`{}\`{}\`{}I find it slightly unsettling to have a reservation at 4:00 and that's when the restaurant opens. Standing outside with several other guest appeared like a soup line during the depression. However a better dressed group. This a bad look. Approximately 10 minutes before the official restaurant opening allow your patrons to enter the lobb foyer area to check in. \textbackslash n\textbackslash nOnce inside everyone was pleasant and professional. I was checked in and seated within 5 minutes. The restaurant is large however I was seated at the opening of the restroom. UGH. Luckily there wasn't much traffic, during Covid you would think more people would be washing their hands, nope!\textbackslash n\textbackslash nI asked for medium well steak however based on the temperature description given by Mindy I was informed the steak would be pink. With more information  I ordered my bone in ribeye medium plus (?).  Although the steak was juicy it was very crisp to 'blackened' and quite pink inside.\textbackslash nNot as described.  The Praline Sweet Potato is listed: whipped - vanilla bean - mascarpone - candied pecans - streusel crisp  it lacked any flavor whatsoever, and was extremely stringy which is a NO\textbackslash nNO! . This dish should be smooth as pudding bursting with flavor and texture. It got better with the king crab meat and avocado stack that was a generous portion with flavor and fresh. The freshly baked hot and slightly buttery rolls were the best item offered during this meal. \textbackslash n\textbackslash nMy server Mindy was professional, prompt and efficient. Perhaps trying to recover for Covid slowness is difficult for some because the meal was a disaster for me. This is no reflection on her Mindy but  I totally wasted \$100.\`{}\`{}\`{}\\
\\
{[\\
    "Opening time is 4:00.",\\
    "Waiting may also be required for an early reservation.",\\
    "Patrons are allowed in approximately 10 minutes before the official restaurant opening.",\\
    "Everyone was pleasant and professional.",\\
    "Checked in and seated is fast.",\\
    "The restaurant is large.",\\
    "Seating position can be bad.",\\
    "Medium well steak would be pink.",\\
    "Bone-in ribeye medium plus is served.",\\
    "Although the bone-in ribeye medium is juicy, it is very crisp to 'blackened' and quite pink inside, not as described.",\\
    "Praline sweet potato lacks flavor and is extremely stringy.",\\
    "King crab meat and avocado stack is a generous portion.",\\
    "King crab meat and avocado stack is flavorful and fresh.",\\
    "Fresh baked hot and slightly buttery rolls are the best item.",\\
    "Server is professional, prompt, and efficient.",\\
    "The meal is a disaster.",\\
    "It feels like a waste of money."
]}\\
\\
Analyze this review: \\
\`{}\`{}\`{}\{review\}\`{}\`{}\`{}
\end{promptbox}

\begin{promptbox}{Prompt for decomposing reviews into item snippets (Amazon) }
\scriptsize
You are a helpful assistant. Follow the instructions. No prose.
\tcblower
\scriptsize
As a language genius, you are tasked with reading book reviews and extracting and summarizing atomic, simple, short and coherent sentences that contain factual descriptions or subjective opinions related to a specific aspect of the book. You should use multiple atomic propositions if the content is about several topics, but combine similar content into one proposition.\\
\\
Ensure these sentences carry information that effectively help differentiate various books. If there is already a proposition with a similar meaning, ignore the redundant information. Ignore the irrelevant chatter, narratives and descriptions unrelated to the properties of the restaurant in the reviews. Try to use original texts from the reviews and but do summarize them if they are verbose. Be sure to cover the whole review. Try to eliminate any references of "I" or "reviewer", but focus on "book". Follow the exemplar format to extract.\\
\\
Analyze this review:\\
\`{}\`{}\`{}I purchased this book when it first came out.  I was initially pleased and made place mats by using one of the patterns. They came out well and I had no problems.  I recently completed a warp to make Serendipity, one of the more attractive patterns featured in the book. I reviewed the directions for threading but counting up the number of ends I would need and comparing the count with the pattern in the book I had a different count. I recounted several times and even tried several ways of adding things up but my end count remained the same. I assume that I misunderstood the sparse directions which were not clear. \`{}\`{}\`{}\\
\\
{[\\
"The book provides patterns for projects like place mats.",\\
"The pattern 'Serendipity' is one of the more attractive options in the book.",\\
"Threading directions were sparse",\\
"Threading directions were unclear",\\
]}\\
\\
Analyze this review:\\
\`{}\`{}\`{}I found it difficult to find the information I needed.  The instructions were not particularly clear to me.\`{}\`{}\`{}\\
\\
{[\\
"It is difficult to find needed information.",\\
"Instructions were not clear."\\
]}\\
\\
Analyze this review:\\
\`{}\`{}\`{}The imagery in this book is so good that I actually slept with my light on a few times. I chose this book because I'm a fan of horror in general and the exorcist is one of my all time favorites. I didn't have high expectations of this book because the subjects been over done but this book reeled me in from the very beginning.\`{}\`{}\`{}\\
\\
{[\\
"The book's imagery is so scary that it caused a need to sleep with the light on.",\\
"The book has connection to the horror genre and similarity to The Exorcist, a classic favorite.",\\
"Despite the overdone subject, the book captivates from the very beginning."\\
]}\\
\\
\\
Analyze this review:\\
\`{}\`{}\`{}Gorgeous coffee table book, chock full of absolutely amazing National Geographic photography. Beautiful landscapes, I particularly loved the northern lights since I am planning a trip to Iceland for 2019. This book would make an absolutely beautiful Christmas, Holiday, birthday, Mother's or Father's Day present for anyone who loves nature I have several NG books and each year they seem to get better and more spectacular!\`{}\`{}\`{}\\
\\
{[\\
"The book is a gorgeous coffee table book.",\\
"The book is filled with amazing National Geographic photography.",\\
"The book features beautiful landscapes.",\\
"The book features the northern lights.",\\
"The book is an ideal gift for nature lovers.",\\
"The book is an ideal gift for occasions like Christmas, holidays, birthdays, Mother's Day, or Father's Day.",\\
]}\\
\\
Analyze this review: \\
\`{}\`{}\`{}\{review\}\`{}\`{}\`{}
\end{promptbox}

\subsection{Query Snippet Extraction (\S~\ref{sec:snippet_user})}
\label{sec:prompt_query_snippet}

\begin{promptbox}{Prompt for decomposing user responses into query snippets (Yelp) }
\scriptsize
You are a helpful assistant.

\tcblower
\scriptsize
You are provided with a pair of a question and a response related to restaurant recommendations. The question asks about preferences for a restaurant, and the response may provide specific information about the user's preferences or indicate that there is no particular preference.\\
\\
**Your task:**\\
\\
1. **Extract Intents:** Identify any stated preferences or dislikes about restaurants in the response.\\
2. **Convert to Requirement Statements:** Write short, complete sentences that objectively describe a requirement for the restaurant search. Avoid using subjective phrases like "the user likes" or "the user wants." Instead, write factual statements.\\
3. **Annotate:**\\
    - \`{}prop\`{}: A brief description of the restaurant feature the user prefers or dislikes.\\
    - \`{}sentiment\`{}: \`{}preference\`{} or \`{}dislike\`{}.\\
\\
**Example 1:**\\
\\
Input:\\
\\
- Question: "What kind of ambiance are you looking for in a restaurant?"\\
- Response: "I want a lively spot with good music and maybe outdoor seating. I'm not into anything too formal."\\
- Known intents: None\\
\\
Output:\\
\\
\`{}\`{}\`{}\\
\lbrack\\
    Intent(prop="the place is lively", sentiment="preference"),\\
    Intent(prop="the place has good music", sentiment="preference"),\\
    Intent(prop="the place has outdoor seating", sentiment="preference"),\\
    Intent(prop="the place is too formal", sentiment="dislike")\\
\rbrack\\
\\
**Example 2:**\\
\\
Input:\\
\\
- Question: "Any preferences for the type of cuisine?"\\
- Response: "I love Burritos!"\\
- Known intents: None\\
\\
Output:\\
\\
\`{}\`{}\`{}\\
\lbrack\\
    Intent(prop="burritos are served", sentiment="preference")\\
\rbrack\\
\`{}\`{}\`{}\\
\\
\\
**Example 3:**\\
\\
Input:\\
\\
- Question: "Do you need parking spots at the restaurant?"\\
- Response: "Parking isn't really an issue for me. I'm just looking for a cozy cafe."\\
- Known intents:\\
    - preference\\
        - this place is a cafe.\\
        - the atmosphere is cozy.\\
\\
Output:\\
\\
\`{}\`{}\`{}\\
\lbrack\rbrack\\
\`{}\`{}\`{}\\
\\
\\
**Example 4:**\\
\\
Input:\\
\\
- Question: "Do you like cafes that are known for their desserts?"\\
- Response: "I'm not specifically looking for desserts."\\
- Known intents: None\\
\\
Output:\\
\`{}\`{}\`{}\\
\lbrack\rbrack\\
\`{}\`{}\`{}\\
\\
**Guidelines:**\\
\\
- Disregard information in the response that doesn't express a preference or dislike.\\
- Annotate only clear and specific intents.\\
- Each intent should address a single topic; separate multiple topics into individual intents.\\
- Known intents (listed in the provided set) should not be repeated. Ensure that any new intent you extract is distinct and does not overlap with the known intents.\\
- If the response is vague or indicates no specific preference (e.g., "I'm open to...", "I'm not specifically looking for..."), return an empty list (`[]`).\\
\\
**Analyze the following question-response pair:**\\
\\
- Question: "\{question\}"\\
- Response: "\{response\}"\\
- Known intents: \{intents\}\\
\\
\end{promptbox}

\begin{promptbox}{Prompt for decomposing user responses into query snippets (Amazon) }
\scriptsize
You are a helpful assistant.

\tcblower
\scriptsize
You are provided with a pair of a question and a response related to book recommendations. The question asks about preferences for a book, and the response may provide specific information about the user's preferences or indicate that there is no particular preference.\\
\\
**Your task:**\\
\\
1. **Extract Intents:** Identify any stated preferences or dislikes about books in the response.\\
2. **Convert to Requirement Statements:** Write short, complete sentences that objectively describe a requirement for the book search. Avoid using subjective phrases like "the user likes" or "the user wants." Instead, write factual statements.\\
3. **Annotate:**\\
    - \`{}prop\`{}: A brief description of the book feature the user prefers or dislikes.\\
    - \`{}sentiment\`{}: \`{}preference\`{} or \`{}dislike\`{}.\\
\\
**Example 1:**\\
\\
Input:\\
\\
- Question: "hat type of characters do you prefer in a book?"\\
- Response: "I enjoy stories with strong female protagonists and well-developed antagonists. I'm not a fan of overly simplistic characters."\\
- Known intents: None\\
\\
Output:\\
\\
\`{}\`{}\`{}\\
\lbrack\\
    Intent(prop="the book has a strong female protagonist", sentiment="preference"),\\
    Intent(prop="the book has a well-developed antagonist", sentiment="preference"),\\
    Intent(prop="the book has overly simplistic characters", sentiment="dislike")\\
\rbrack\\
\\
**Example 2:**\\
\\
Input:\\
\\
- Question: "Is there a specific genre you're interested in?"\\
- Response: "I love mystery novels!"\\
- Known intents: None\\
\\
Output:\\
\\
\`{}\`{}\`{}\\
\lbrack\\
    Intent(prop="the genre is mystery", sentiment="preference")\\
\rbrack\\
\`{}\`{}\`{}\\
\\
\\
**Example 3:**\\
\\
Input:\\
\\
- Question: "Do themes like adventure or romance matter to you in a book?"\\
- Response: "Themes aren't that important to me. I'm just looking for a good biography."\\
- Known intents:\\
    - preference\\
        - the book is a biography.\\
\\
Output:\\
\\
\`{}\`{}\`{}\\
\lbrack\rbrack\\
\`{}\`{}\`{}\\
\\
\\
**Example 4:**\\
\\
Input:\\
\\
- Question: "Do you prefer books with extensive world-building?"\\
- Response: "I'm not specifically looking for books with elaborate settings."\\
- Known intents: None\\
\\
Output:\\
\`{}\`{}\`{}\\
\lbrack\rbrack\\
\`{}\`{}\`{}\\
\\
**Guidelines:**\\
\\
- Disregard information in the response that doesn't express a preference or dislike.\\
- Annotate only clear and specific intents.\\
- Each intent should address a single topic; separate multiple topics into individual intents.\\
- Known intents (listed in the provided set) should not be repeated. Ensure that any new intent you extract is distinct and does not overlap with the known intents.\\
- If the response is vague or indicates no specific preference (e.g., "I'm open to...", "I'm not specifically looking for..."), return an empty list (`[]`).\\
\\
**Analyze the following question-response pair:**\\
\\
- Question: "\{question\}"\\
- Response: "\{response\}"\\
- Known intents: \{intents\}\\
\\
\end{promptbox}


\subsection{Query Snippet Expansion (\S~\ref{sec:snippet_user})}
\label{sec:prompt_query_expansion}

\begin{promptbox}{Prompt for paraphrasing snippets}
\scriptsize
You are a helpful assistant. Strictly follow the format of the examples; do not provide anything other than the direct answer.

\tcblower
\scriptsize
Paraphrase a given sentence.\\
The sentence should be an atomic, simple, short and coherent sentence that contain factual descriptions or subjective opinions related to a specific aspect of the restaurant/book.\\
\\
Paraphrase this sentence: "\{sentence\}"
\end{promptbox}

\begin{promptbox}{Prompt for generating supporting snippets}
\scriptsize
You are a helpful assistant. Strictly follow the format of the examples; do not provide anything other than the direct answer.

\tcblower
\scriptsize
(Yelp)\\
\\
Generate a sentence that could serve as evidence for a given sentence.\\
The sentence should be an atomic, simple, short and coherent sentence that contain factual descriptions or subjective opinions related to a specific aspect of the restaurant.\\

<example>\\
Given sentence: "the restaurant is located in a bad neighborhood"\\
the restaurant is near bad crime area.\\
\\
Given sentence: "the place is vegetarian-friendly."\\
the menu contains some veggie options.\\
\\
Given sentence: "the place is good for family dinner."\\
high chairs are available.\\
<\textbackslash example>\\
\\
(Amazon)\\
\\
Generate a sentence that could serve as evidence for a given sentence.\\
The sentence should be an atomic, simple, short and coherent sentence that contain factual descriptions or subjective opinions related to a specific aspect of the book.\\
\\
<example> \\
Given sentence: "The book is a thrilling mystery." \\
The plot involves unexpected twists. \\
\\
Given sentence: "The book is suitable for children."\\
There are colorful illustrations.\\
\\
Given sentence: "The book is well-researched."\\
It includes detailed historical references.\\
<\textbackslash example>
\\
Given sentence: "\{sentence\}"

\end{promptbox}

\begin{promptbox}{Prompt for generating opposing snippets}
\scriptsize
You are a helpful assistant. Strictly follow the format of the examples; do not provide anything other than the direct answer.

\tcblower
\scriptsize
(Yelp)\\
\\
Generate the sentence of opposite meaning in restaurant domain following the examples.\\
\\
<example>\\
What's the opposite of this sentence: "this place has sweet options like Cannolis"\\
This place lacks sweet options like Cannolis.\\
\\
What's the opposite of this sentence: "this place is too crowded."\\
This place is very spacious.\\
\\
(Amazon) \\
\\
Generate the sentence of opposite meaning in book domain following the examples.\\
\\
<example> \\
What's the opposite of this sentence: "This book has thrilling moments like a detective novel."\\
This book lacks thrilling moments like a detective novel.\\
\\
What's the opposite of this sentence: "This book is very predictable."\\
This book is full of surprises.\\
<\textbackslash example>

What's the opposite of this sentence: "\{sentence\}"
\end{promptbox}


\subsection{Clarification Question Generation (\S~\ref{sec:cq})}
\label{sec:prompt_CQ}

\begin{promptbox}{Prompt for generating clarification questions}
\scriptsize
You are a Recommender chatting with a Seeker to provide restaurant/book recommendation. Your task is to ask questions for understanding the Seeker's preference.

\tcblower
\scriptsize
\# Role-Play Task: Recommender\\
You will play the role of a Recommender helping a Seeker find a restaurant/book that suits the Seeker's preferences.\\
\\
Based on the conversation log provided below, identify the most relevant aspect of the Seeker's preferences that will help refine the search for a suitable restaurant/book. Your question should focus only on one topic. Do not ask about multiple topics at once.\\
\\
**Dialogue Context:**\\
\{context\}\\
Recommender:\\
\\
Now, generate a response in the role of the Recommender.\\
\\
**Response Guidelines:**\\
\\
- Your response should be concise, typically one sentence. Avoid asking multiple questions at once.\\
- Do not ask for a restaurant/book name or any personal or street names.\\
- Respond directly and concisely to the scenario without repeating the instructions or adding unrelated details. Use question types that give the human user more flexibility, allowing for creative and open-ended answers while staying relevant to the context.
\end{promptbox}


\subsection{User Simulation (\S~\ref{sec:user-simulator})}
\label{sec:prompt_simulator}

\begin{promptbox}{Prompt for user simulators}
\scriptsize
You are a Seeker who is looking for a restaurant/book recommendation. No prose, be concise and casual in the conversation.

\tcblower
\scriptsize
\# Role-Play Task: Seeker\\
You will play the role of a Seeker looking for a restaurant/book recommendation. You will interact with a Recommender to find a restaurant/book that suits your preferences.\\
\\
**Instructions:**\\
\\
- The Recommender will ask for your preferences to identify a restaurant/book that aligns with your tastes. Your role is to provide responses as hints based on the details below.\\
- Express your preferences by answering the Recommender's questions based on the information provided below.\\
\\
**Details of Your Favorite Restaurant/Book:**\\
\\
\{item info\}\\
\{item review summary\}\\
\\
**Your Opinion About It:**\\
\\
\`{}\`{}\`{}\\
\{review text\}\\
\`{}\`{}\`{}\\
\\
**Dialogue Context:**\\
\\
\{dialogue context\}\\
Seeker:\\
\\
Now, generate a response in the role of the Seeker based on the information provided above.\\
\\
**Response Guidelines:**\\
\\
- Your response should be concise, typically one sentence. Avoid giving multiple preference details at once.\\
- If the data provided lacks specifics to answer the Recommender, communicate no particular preference. Try not to invent details.\\
- Tailor your responses around `Your Opinion About It` rather than `Details of Your Favorite Restaurant/Book` when there is conflicting information.\\
- Keep your answers limited to the question asked.\\
- Do not reveal the name of your favorite restaurant/book or any personal or street names.\\
- Focus on answering the Recommender's questions. Do not proactively ask questions such as, "What kinds of restaurants/books are there?" or "Can you tell me about different cuisines in the area?"
\end{promptbox}



\section{User Simulation Example}
\label{sec:appendix_simulator}
We present an example of user simulation. Below are the selected item review, attribute information, and summary of its other favorable reviews, which are used to build the context provided to the simulator as in Appendix~\ref{sec:prompt_simulator}.

\begin{promptbox}{Example context provided to user simulator prompt (\S~\ref{sec:prompt_simulator})}
\scriptsize
\tcbsubtitle{\{review text\}}
Tried the cart at a street. I ordered the General Tso's chicken stir fry. I talked to our server about how spicy and he was like it's not that spicy but it definitely has a kick (I'm also a wimp haha) but the flavor is great. The guy working was super nice and personable while helping me order. The veggies taste fresh and are crisp and I am very happy with the size of the portion for \$10. I'll definitely have a bit for leftovers!"

\tcbsubtitle{\{item info\}}
Category: asian fusion\\
- Alcohol: none\\
- Ambience/touristy: False\\
- Ambience/hipster: False\\
- Ambience/romantic: False\\
- Ambience/divey: False\\
- Ambience/intimate: False\\
- Ambience/trendy: False\\
- Ambience/upscale: False\\
- Ambience/classy: False\\
- Ambience/casual: False\\
- BYOB: No\\
- BikeParking: No\\
- BusinessAcceptsBitcoin: Yes\\
- BusinessAcceptsCreditCards: Yes\\
- BusinessParking: None\\
- Caters: Yes\\
- DogsAllowed: Yes\\
- GoodForKids: No\\
- GoodForMeal/dessert: False\\
- GoodForMeal/latenight: False\\
- GoodForMeal/lunch: False\\
- GoodForMeal/dinner: False\\
- GoodForMeal/brunch: False\\
- GoodForMeal/breakfast: False\\
- HappyHour: No\\
- HasTV: No\\
- OutdoorSeating: Yes\\
- RestaurantsDelivery: Yes\\
- RestaurantsGoodForGroups: No\\
- Price Range: $11-$30\\
- RestaurantsReservations: No\\
- RestaurantsTableService: No\\
- RestaurantsTakeOut: Yes\\
- WheelchairAccessible: No\\
- WiFi: no

\tcbsubtitle{\{item review summary\}}
What people generally like: Customers appreciate the delicious food, particularly the drunken noodles and chili garlic sauce, which many found flavorful and well-prepared. The freshness of the vegetables and the generous portion sizes receive high praise, with some meals lasting enough for two servings. The friendly and helpful staff enhance the dining experience, creating a positive atmosphere at this restaurant. Convenient location adds to the appeal, making it a popular lunch spot for locals. Many reviewers recommend trying the different combinations and sauces available.	    
\end{promptbox}

\begin{promptbox}{Example context provided to user simulator prompt (\S~\ref{sec:prompt_simulator})}
\scriptsize
\tcbsubtitle{\{review text\}}
The author knows of what he writes. The book, while a relatively easy read, backs up every allegation with sources and endnotes that in of themselves are worth reading and pursuing.<br /><br />Anyone concerned about "abuse of power" and "imperial presidency" needs to at least seriously read the book and - better yet - follow up on the ample source materials found in the endnotes (and while some are web sites, the articles in the sites are based on fact and not opinion. But more often than not, you get actual government, congressional documents).<br /><br />I find it amazing how any book that isn't pro-Administration pablum, especially those written by people in the know, immediately attracts the usual suspects from the right-wing ranting over how bad the book is - and of course give little evidence of having read it, or else concentrate on one or two little items (i.e. was a certain official dissimulating his health problems?) that mask the very very serious topics the book examines.<br /><br />Why is this current Administration worse than Nixon's. For starters, while Nixon used/abused all the tools: executive privilege, the courts, stonewalling, hyper-secrecy etc to try and save a floundering Administration, the current Administration are using and abusing them to establish new and dangerous definitions of presidential power for its own sake.<br /><br />And all this without even a word about how the current Republican controlled Congress is in its own arrogant way redefining the "majority rules" iron fist over the minority.

\tcbsubtitle{\{item info\}}
- Category: Books > Politics \& Social Sciences > Politics \& Government\\
- Author: John W. Dean\\
- Features: The former cousel to President Nixon provides a stinging critique of the current Bush administration, its obsession with secrecy, and its willingness to deceive the American people, emphasizing the president's emphasis on image over substance, mistrustful personality, imperial governing and flawed decision making, and his abuses of national security secrecy. 75,000 first printing.\\
- Description: Amazon.com Review
    The most facile presidential comparison one could make for George W. Bush would be his father, who presided over a war in Iraq and a struggling economy. Some ""neocons"" reject the parallel and compare Bush to his father's predecessor, Ronald Reagan, citing a plainspoken quality and a belief in deep tax cuts. But John Dean goes further back, seeing in Bush all the secrecy and scandal of Dean's former boss, the notorious Richard Nixon. The difference, as the title of Dean's book indicates, is that Bush is a heck of a lot worse. While the book provides insightful snippets of the way Nixon used to do business, it offers them to shed light on the practices of Bush. In Dean's estimation, the secrecy with which Bush and Dick Cheney govern is not merely a preferred system of management but an obsessive strategy meant to conceal a deeply troubling agenda of corporate favoritism and a dramatic growth in unchecked power for the executive branch that put at risk the lives of American citizens, civil liberties, and the Constitution. Dean sets out to make his point by drawing attention to several areas about which Bush and Cheney have been tight-lipped: the revealing by a ""senior White House official"" of the identity of an undercover CIA operative whose husband questioned the administration, the health of Cheney, the identity of Cheney's energy task force, the information requested by the bi-partisan 9/11 commission, Bush's business dealings early in his career, the creation of a ""shadow government"", wartime prisoners held at Guantanamo Bay, and scores more. He theorizes that the truth about these and many other situations, including the decision to go to war in Iraq, will eventually surface and that Bush and Cheney's secrecy is a thus far effective means of keep a lid on a rapidly multiplying set of lies and scandals that far outstrip the misdeeds that led directly to Dean's former employer resigning in disgrace. Dean's charges are impassioned and more severe than many of Bush's most persistent critics. But those charges are realized only after careful reasoning and steady logic by a man who knows his way around scandal and corruption.
    --John Moe
    From Publishers Weekly
    This title's accusation bears particular weight coming from the man who warned the super-secretive Richard Nixon that there was a cancer on his presidency, and Dean, who was Nixons White House counsel, makes a strong argument that the secrecy of what he dubs the ""Bush-Cheney presidency"" is ""not merely unjustified and excessive but obsessive,"" and consequently ""frighteningly dangerous."" Some of the subjects he touches on have been covered in detail elsewhere, and his chapter on the administration's stonewalling of the September 11 commission isn't fully up to date. But few critics have as effectively put the disparate pieces together, linking them to what Dean says is a broader pattern of secrecy from an administration that does its best to control the flow of information on every subject even the vice presidents health and uses executive privilege to circumvent congressional scrutiny. Deans probe extends back to Bushs pre-presidential activities, such as his attempt to withhold his gubernatorial papers from public view, and Deans background as an investment banker adds welcome perspective on Bushs business career (as well as Cheneys). Dean ultimately identifies 11 issues (such as the secrecy around the forming of a national energy policy and what Dean calls Bushs misleading of Congress about war with Iraq) on which the White Houses stance could lead to scandal, and warns that allowing the administration to continue its policy of secrecy may lead to a weakening of democracy. Despite occasional comments about Bushs intelligence that will rankle presidential supporters, Dean (Blind Ambition) is generally levelheaded; his role in Watergate and the seriousness of his charge in the national media that Bush has committed impeachable offenses has popped this onto bestseller lists. Copyright Reed Business Information, a division of Reed Elsevier Inc. All rights reserved.
    From the Publisher
    Former White House counsel and New York Times bestselling author John Dean reveals how the Bush White House has set America back decades-employing a worldview and tactics of deception that will do more damage to the nation than Nixon at his worst.
    About the Author
    John W Dean's previous works include Blind Ambition, Lost Honor, The Rehnquist Choice, and Unmasking Deep Throat.
    Read more\\
- Price: \$None"

\tcbsubtitle{\{item review summary\}}
- What people generally like: Readers appreciated the well-researched and organized critique of the Bush administration, considering it a credible and authoritative examination due to the author's background as a former White House counsel. Many found the book engaging and compelling, highlighting its thorough analysis of secrecy and its implications for democracy. Reviewers noted the similar historical context drawn from the Nixon era, recognizing the author's expertise in linking past and present political scandals. Some readers felt the book was important for understanding the motivations behind major decisions, such as the Iraq war, and urged others to read it for a deeper perspective. Overall, the book resonated well with those critical of Bush's presidency, prompting discussions about the need for accountability in government.    
\end{promptbox}

Using the above exemplar item, we also provide sample ratings for selected user responses on three dimensions: relevancy, on-topicness, faithfulness.

\begin{itemize}
\setlength\itemsep{-0.5em}
    \item \textbf{Question:} What type of atmosphere are you hoping for, like casual, upscale, or something in between?
    \item \textbf{Answer:} I'm looking for something casual.
    \item \textbf{Rating:} Relevant $\checkmark$, On-topic $\checkmark$, Faithful $\times$
\vspace{.5em}
    \item \textbf{Question:} What specific dishes or flavors are you most interested in trying at an Asian fusion restaurant?
    \item \textbf{Answer:} I'm interested in trying dishes like drunken noodles and anything with chili garlic sauce.
    \item \textbf{Rating:} Relevant $\checkmark$, On-topic $\checkmark$, Faithful $\checkmark$
\vspace{.5em}
    \item \textbf{Question:} Are you looking for a specific type of cuisine within Asian fusion, like Thai, Chinese, or Japanese influences?
    \item \textbf{Answer:} I'm open to any influences, but I really enjoyed the General Tso's chicken stir fry.
    \item \textbf{Rating:} Relevant $\times$, On-topic $\times$, Faithful $\times$
\end{itemize}



\section{Evaluation of Item Snippets}

\label{item-snippet-hallucination}

We used LLM-as-judge to ask gpt-4o to judge whether a snippet can be inferred from a paragraph. We sampled 500 snippets from the decomposition results of each dataset. Among these, the proportion of snippets containing hallucinations was 2.49\% in the restaurant dataset, 3.90\% in the book dataset, and 2.90\% in the clothing dataset. The prompt used for the LLM-as-judge is as follows:

\begin{promptbox}{Prompt for Judging Hallucination}
\scriptsize
Work as a judge and determine if the following proposition can be inferred from the given customer review in the {keywords}. No prose.

\tcblower
\scriptsize
Can the following proposition be inferred from the given customer review in the restaurant domain?\\
\\
Proposition: {hypo}\\
Customer Review: {premise}\\
\\
Answer with 'yes' or 'no'.
\end{promptbox}

To further examine the reliability of this LLM-as-judge, two of the authors manually annotated 100 snippets, 3/100 are deemed as hallucinations. Overall, we found that our extractor is faithful to the given context. In our manual assessment, we also evaluated atomicity using the following definition: A sentence $(s)$ is not atomic if it can be decomposed into a conjunction of shorter sentences $(s_1)$, $(s_2)$, etc., such that any combination of these sentences is entailed by the original sentence. In other words, $(s)$ entails $(s_1)$, $(s)$ entails $(s_2)$, and so on. If these conditions are not met, then $(s)$ is considered atomic. We apply this criterion to two levels: the aspect level (e.g., food, location) and the proposition level. The proposition-level atomicity corresponds to the decompositions considered in existing studies~\cite{wanner2024closer}. As noted in Section~\ref{sec:snippet_evaluation}, while snippets are often not atomic at the proposition level, they are generally atomic at the aspect level.


Table \ref{tab:hallucination_snippets} are some examples.

    \begin{table*}[h]
        \rotatebox[]{90}{
        \centering
        \renewcommand{\arraystretch}{1.2}
        \setlength{\tabcolsep}{4pt} 
        \small 
        \begin{tabular}{|c|p{2cm}|p{8.5cm}|c|c|c|}
            \hline
            \textbf{Category} & \textbf{Extracted Snippet} & \textbf{Original Review} & \textbf{Atomic Aspect} & \textbf{Atomic Proposition} & \textbf{Hallucination} \\
            \hline
            \multirow{2}{*}{Hallucination Examples} 
            & Discount is offered on bulk purchases. 
            & This place has such an amazing beer selection! I cleaned them out of their Southern Tier Pumpking the other night. I ended up buying their last four bottles, so they hooked me up with a 10\% discount. The staff here is very friendly and knowledgeable. I love bringing my family rare beers to share during the holiday season, so I'll definitely be back to see what else I can get my hands on!
            & Yes & No & Yes \\
            \cline{2-6}
            & The corner store is the best in the area. 
            & I love this place. I give their service and their tofu hoagie two enthusiastic thumbs up! Not only that, but they have a fairly good selection of groceries most corner stores don't: rice noodles, vegetarian meals, dried goods, etc... and as long as you call ahead, there's usually no wait for prepared foods. (I recommend calling ahead for hoagies, especially on weekends, when the place is jumpin'.) Fu-Wah is the best corner store in the area, at least in my opinion.
            & Yes & No & Yes \\
            \hline
            \multirow{2}{*}{Good Atomic Snippets} 
            & The book is highly recommended for GRE preparation. 
            & I had to take the GRE in two weeks because I decided at the last second to apply for business school. Highly recommended! If you're good at math, but not excellent or needed a refresher, you can skim the pages and do some exercises. Then take the practice tests and go over what you did wrong - fast and easy way to refresh your memory and excel on the test!
            & Yes & Yes & No \\
            \cline{2-6}
            & The book provides explanations for recipe components. 
            & It's a great cookbook with good recipes. But we enjoyed the "why" in the descriptions. For example, why did it need another egg yolk. So you'll actually learn more about why recipes are constructed like they are.
            & Yes & Yes & No \\
            \hline
            \multirow{2}{*}{Atomic Aspect but Further Decomposable} 
            & This place is for the lovers and the dreamers. 
            & I walk into the tune of Regulate by Nate Dogg and Warren G. The Godfather part 2 is playing and the place screams ambiance. It's like this place knows me. >> Rene is sitting by the door, on his phone. I've never actually met the guy, but he's legendary and so his reputation precedes him. There's not much you are going to learn from my review that you won't get from everyone else's but if you haven't gone, you should go. While you're at it, bring a friend because this place is for the lovers, the dreamers, and me.
            & Yes & No & No \\
            \cline{2-6}
            & The menu fills the wall behind you and is made with individual letter pins. 
            & At first I balked at the idea of spending \$11 for a sandwich, but this place was highly recommended by local friends. I stopped by here for lunch and ordered the Shopsin Club, and it is simply amazing. The store is a small shop with a counter and a few tables, and the cashier warmly greets you when you enter. >> The menu fills the wall behind you and is made with individual letter pins. The sandwich itself is perfect in every way -- crisp toasted ciabatta bread, thick sliced turkey, avocado, bacon, mayo, pickles. I contemplated leaving out the pickles, but was glad I kept it as is as it really goes well with the rest of the sandwich. Everything is lovingly made to order and carefully wrapped in paper and sliced.
            & Yes & No & No \\
            \hline
        \end{tabular}
        }
        \caption{Hallucination and Atomicity Analysis of Snippets}
        \label{tab:hallucination_snippets}
    \end{table*}

\section{Conversational Recommendation Examples}
\label{sec:interaction}

Tables~\ref{tab:interactions1} and \ref{tab:interactions2} contain 2 cases, each with 5 turns. 

\begin{table*}[h]
\small
\centering
\rotatebox[]{90}{
\renewcommand{\arraystretch}{1.5}
\begin{tabular}{|p{4cm}|p{2cm}|p{1.8cm}| p{3cm}|c|p{3cm}|p{1.8cm}|p{3cm}|c|}
\hline
&\multicolumn{4}{|c|}{\textbf{GPT-4o-mini}} & \multicolumn{4}{|c|}{\textbf{LLaMa-3.3 70B}}\\
\cline{2-9}
\textbf{Context} & \textbf{Question} & \textbf{Answer} & \textbf{Snippets} &\textbf{Rank} & \textbf{Question} & \textbf{Answer} & \textbf{Snippets} & \textbf{Rank}\\ \hline
\multirow{5}{*}{\parbox{4cm}{
\vspace{.5\baselineskip}
\tiny 
I love this restaurant, so if you are a tourist, this is a solid option for a real cheesesteak in this city! Also they have amazing fries, get some cheese on them as well. As for us, we have hit up this restaurant in various locations but this one was the best because it saved us from the frigid cold of the Christmas Village at a park. There are not too many places to eat near that park and I was craving a cheesesteak so this was perfect. This location has a decent amount of seating area so that's a huge plus as well. We came with our 1 year old who officially had his first bite of a cheesesteak at this restaurant! All in all, would absolutely come back.	\\

- Category: cheesesteaks
- Alcohol: none
- Ambience/touristy: None
- Ambience/hipster: False
- Ambience/romantic: False
- Ambience/divey: None
- Ambience/intimate: False
- Ambience/trendy: False
- Ambience/upscale: False
- Ambience/classy: False
- Ambience/casual: True
- BikeParking: Yes
- BusinessAcceptsBitcoin: No
- BusinessAcceptsCreditCards: Yes
- BusinessParking/garage: False
- BusinessParking/street: True
- BusinessParking/validated: False
- BusinessParking/lot: False
- BusinessParking/valet: False
- Caters: Yes
- DogsAllowed: No
- DriveThru: No
- GoodForKids: Yes
- GoodForMeal/dessert: False
- GoodForMeal/latenight: True
- GoodForMeal/lunch: True
- GoodForMeal/dinner: True
- GoodForMeal/brunch: False
- GoodForMeal/breakfast: False
- HappyHour: No
- HasTV: Yes
- NoiseLevel: average
- OutdoorSeating: No
- RestaurantsAttire: casual
- RestaurantsDelivery: Yes
- RestaurantsGoodForGroups: Yes
- Price Range: \$11-\$30
- RestaurantsReservations: No
- RestaurantsTableService: No
- RestaurantsTakeOut: Yes
- WheelchairAccessible: Yes
- WiFi: no
\\
- What people generally like: Many reviewers enjoyed the cheesesteaks, noting the soft bread and the tenderness of the steak. Some described the portion sizes as generous and appreciated the variety of accompaniments available. Positive mentions were made about the taste of the wiz fries and the overall flavor of the sandwiches. A few customers praised the quick service and the laid-back environment of this restaurant. This restaurant is seen as a classic spot for cheesesteaks in a city known for them.}} & Hello, what category of restaurant are you looking for? & I'm looking for a place that serves cheesesteaks. & {\tiny \{'preference': ['cheesesteaks are served', 'the menu features a variety of cheesesteak sandwiches.', '"Cheesesteaks are offered."'], 'dislike': ['Cheesesteaks are not served.']\} }& 11 & Hello, what category of restaurant are you looking for? & I'm looking for a place that serves cheesesteaks. & {\tiny \{preference': ['the place serves cheesesteaks', 'the menu includes a philly cheesesteak sandwich.', 'They offer cheesesteaks here.'], 'dislike': ['The place does not serve cheesesteaks.']\}} & 13\\ 

\cline{2-9}
 & What type of atmosphere do you prefer for your cheesesteak experience, casual or more upscale? & Casual is definitely my vibe for cheesesteaks.& { \tiny \{'preference': ['the atmosphere is casual', 'customers often wear jeans and t-shirts.', 'The vibe is relaxed.'], 'dislike': ['The atmosphere is formal.'\}}& 10 & What kind of atmosphere are you envisioning for your cheesesteak dining experience? & I'm looking for a casual atmosphere.& { \tiny \{'preference': ['the place has a casual atmosphere', 'people wear jeans and t-shirts when dining there.', 'The restaurant has a relaxed ambiance.'], 'dislike': ['The place has a formal atmosphere.']\}} & 11\\ \cline{2-9}
 & What is your preferred location or neighborhood for enjoying your casual cheesesteak? & I'm open to any location, just looking for a good cheesesteak spot. & { \tiny \{\}}& 10 & How important is it for you to have a wide variety of cheesesteak options, such as different types of cheese or meats, at the restaurant? & I don't have a specific preference for the variety of cheesesteak options, just a solid cheesesteak is good for me! & { \tiny \{\}} & 11\\ \cline{2-9}
 & What is your preferred price range for a casual cheesesteak? & I'm looking for a price range between \$11 and \$30 for a cheesesteak. & { \tiny \{'preference': ['the price range is between \$11 and \$30', 'most main dishes cost around \$20.', 'The prices vary from \$11 to \$30.'], 'dislike': ['The price range is not between \$11 and \$30.']\}}& 7 & What is your preferred price range for this casual cheesesteak dining experience? & I'm looking for something in the \$11-\$30 range. & { \tiny \{'preference': ['the price range is between \$11 and \$30', 'most meals cost around twenty dollars.', '"Prices fall within the \$11 to \$30 range"'], 'dislike': ['The price range is outside of \$11 and \$30, or more specifically: The price range is either under \$11 or over \$30.']\}}& 2\\ \cline{2-9}
 & What specific toppings or ingredients do you enjoy on your cheesesteak? & I enjoy classic toppings, but I'm not picky about specifics. &{ \tiny \{'preference': ['classic toppings are enjoyed', 'pepperoni and mushrooms are popular choices.', '"Traditional toppings are appreciated."'], 'dislike': ['Classic toppings are disliked.']\}}& 0 & Are you open to trying a restaurant that is a bit of a walk from public transportation, or would you prefer one that is more conveniently located near a bus or train stop? & I'm okay with a bit of a walk from public transportation. & { \tiny \{'preference': ['the place is within walking distance from public transportation', 'a bus stop is right in front of the restaurant.', 'The restaurant is easily accessible on foot from the nearest public transport stop.'], 'dislike': ['The place is far from public transportation.']\}}& 7\\ \hline
\end{tabular}}
\caption{User and \system{} + Expansion interaction excerpts from Yelp dataset. Context is provided to the user simulator, and rank is the current position of the target item among all candidates.}
\label{tab:interactions1}
\end{table*}

\begin{table*}[h]
\tiny
\centering
\rotatebox[]{90}{
\renewcommand{\arraystretch}{1.5}
\begin{tabular}{|p{4cm}|p{1.8cm}|p{1.4cm}| p{4cm}|c|p{2.5cm}|p{1.4cm}|p{4cm}|c|}
\hline
&\multicolumn{4}{|c|}{\textbf{GPT-4o-mini}} & \multicolumn{4}{|c|}{\textbf{LLaMa-3.3 70B}}\\
\cline{2-9}
\textbf{Context} & \textbf{Question} & \textbf{Answer} & \textbf{Snippets} &\textbf{Rank} & \textbf{Question} & \textbf{Answer} & \textbf{Snippets} & \textbf{Rank}\\ \hline
\multirow{5}{*}{\parbox{4cm}{
\vspace{.5\baselineskip}
\tiny 
This takes some liberty with the historical facts. For example, Queen Mary quite approved of Philip's marriage to Elizabeth; King George VI never knew that he had lung cancer. Jared Harris does not look like King George VI, but he does a splendid job of portraying the monarch. What the book lacks in historical accuracy, it captures the spirit of the events (even when that spirit dissolves into fiction such as the "fog" and the death of Churchill's Venetia who never did exist. Some of my favorite moments were the Queen Mother acquiring the Castle of Mey, Queen Mary's stern advice to the new Queen given at the bedside of the old Queen, and almost anything that John Lithgow's Churchill utters. I loved his comments about Philip's sisters, Mountbatten, and Clement Attlee. I cannot understand the fictional nature of some of the scenes such as the scene with Philip and the bull elephant, the wedding scene which looked nothing like Westminster Abbey, even though I did like Churchill's comment about why Philip's sisters were not present, and I laughed out loud at Queen Elizabeth (mother of the Princess) and her comment about Philip's mother. There is much to be enjoyed here, and I look forward to the time when the series will be available on DVD.
- Category: Books > Biographies \& Memoirs > Arts \& Literature
- Author: Robert Lacey
- Features: The official companion to the Emmy-winning Netflix drama chronicling the reign of Queen Elizabeth II, and starring Claire Foy and John Lithgow, The Crown by Peter Morgan, featuring additional historical background and beautifully reproduced archival photos and show stills Elizabeth Mountbatten never expected her father to die so suddenly, so young, leaving her with a throne to fill and a global institution to govern. Crowned at twenty-five, she was already a wife and mother as she began her journey towards becoming a queen... Princess Elizabeth had come of age. She was ready and she needed to be. When George VI returned from his 1947 tour of South Africa, he was looking and feeling dreadful. In his twelve weeks away he had lost 17lb in weight.
    Read more
- Price: \$10.29
\\
- What people generally like: Readers appreciated the book as a great companion to the series, providing historical details that enrich the context of the drama. The layout and quality of images from the show were noted as visually impressive, and many found the historical insights into royal life fascinating. Fans of the series found the book to clarify the dramatized events and enhance their understanding of the characters and plot. The author's writing style was praised for being interesting and concise, making the historical narrative engaging. Overall, it was recommended for both fans of the series and those interested in royal history.}} & Hello, which category of books are you looking for? & I'm looking for a biography or memoir, especially related to arts and literature. & {\tiny \{'preference': ['It chronicles the life of a famous individual.', 'It discusses various literary movements.', 'It features discussions on famous artists.', "It recounts personal experiences from the author's life.", 'The book is a life story.', 'The book is a personal narrative.', 'The book pertains to literature.', 'The book pertains to the arts.', 'the book is a biography', 'the book is a memoir', 'the book is related to arts', 'the book is related to literature'], 'dislike': ['The book is a work of fiction.', 'The book is unrelated to literature.', 'The book is unrelated to the arts.']\} }& 7038 & Hello, which category of books are you looking for? & I'm looking for biographies or memoirs, especially related to arts and literature. & {\tiny \{{'preference': ['the book is a biography', 'the book is a memoir', 'the book is related to arts and literature', "It describes the subject's life events in chronological order.", 'The author recounts their personal experiences.', 'It discusses various famous authors and their works.', "The book tells the story of someone's life.", 'The book is an autobiography.', 'The book focuses on creative fields such as art and literature.'], 'dislike': ['The book is not a biography.', 'The book is not a memoir.', 'The book is unrelated to arts and literature.']}\}} & 203\\ 

\cline{2-9}
 & What specific time period or artist's life are you most interested in exploring through a biography or memoir? & I'm interested in biographies that capture the spirit of historical events, even if they take some liberties with the facts. & { \tiny \{{'preference': ['It vividly describes key moments from the past.', 'The author alters historical events for dramatic effect.', 'The book alters some of the facts.', 'The book reflects the essence of historical events.', 'the book captures the spirit of historical events', 'the book takes some liberties with the facts'], 'dislike': ['The book adheres strictly to the facts.', 'The book fails to convey the essence of historical events.']}\}}& 8743 & What specific aspects of arts and literature, such as the life of a particular artist, writer, or movement, interest you the most in a biography or memoir? & I'm really into royal history and the personal lives of historical figures. & { \tiny \{{'preference': ['the book is about royal history', 'the book is about the personal lives of historical figures', 'It describes the lives of several monarchs.', 'It describes the daily routines of famous leaders.', 'The book focuses on the history of royalty.', 'The book explores the private lives of famous historical individuals.'], 'dislike': ['The book is not about royal history.', 'The book is not about the personal lives of historical figures.']}\}} & 60\\ \cline{2-9}
 & What specific historical events or figures in the arts and literature are you particularly drawn to? & I'm drawn to biographies that explore royal history and the complexities of personal relationships, even if they aren't entirely factually accurate. & { \tiny \{{'preference': ['It details the lives of several monarchs.', 'The book examines the history of royalty.', 'The book examines the intricacies of personal connections.', 'The characters experience deep emotional conflicts.', 'the book explores royal history', 'the book explores the complexities of personal relationships'], 'dislike': ['The book avoids the intricacies of personal relationships.', 'The book ignores royal history.']}\}}& 242 & What time period or historical era would you prefer the royal history or personal lives of historical figures to be from in the biography or memoir you're looking for? & I'm particularly interested in the 20th century, especially the British royal family. & { \tiny \{{'preference': ['the book is set in the 20th century', 'the book is about the British royal family', 'The story takes place during World War II.', 'The story revolves around the lives of kings and queens of England.', 'The story takes place during the 20th century.', 'The book focuses on the British monarchy.'], 'dislike': ['The book is not set in the 20th century.', 'The book is not about the British royal family.']}\}} & 20\\ \cline{2-9}
 & What specific royal figures or historical events related to the arts and literature are you most interested in? & I'm particularly interested in biographies that delve into royal figures and their personal relationships, even if they take some creative liberties with the facts. & { \tiny \{{'preference': ['It explores the lives of several kings and queens.', 'The book examines individual connections.', 'The book explores royal characters.', 'The characters have deep emotional connections.', 'the book delves into royal figures', 'the book explores personal relationships'], 'dislike': ['The book avoids any mention of royal figures.', 'The book ignores personal relationships.']}\}}& 70 & What kind of experiences or aspects of the British royal family's lives in the 20th century would you like the biography or memoir to focus on, such as their struggles, achievements, or personal relationships? & I'm interested in their struggles and personal relationships, especially how they navigated challenges during that time. & { \tiny \{'preference': ['the book focuses on the struggles of the British royal family', 'the book focuses on the personal relationships of the British royal family', 'The story revolves around the personal lives of kings and queens.', 'The story explores the intimate interactions between Queen Elizabeth and her siblings.', 'The book explores the challenges faced by the British monarchy.', 'The book explores the interpersonal dynamics within the British monarchy.'], 'dislike': ['The book focuses on something other than the struggles of the British royal family.', 'The book ignores the personal relationships of the British royal family.']\}} & 27\\ \cline{2-9}
 & Which specific royal figures or dynasties are you most fascinated by in terms of their personal relationships and artistic contributions? & I'm really fascinated by Queen Elizabeth II and her relationships, especially how they played out against the backdrop of royal duties. &{ \tiny \{{'preference': ['It explores her reign and significant events.', 'It provides insights into her friendships and rivalries.', "The book examines Queen Elizabeth II's connections.", 'The book focuses on Queen Elizabeth II.', 'The characters navigate court politics.', 'The story takes place amid royal responsibilities.', "the book explores Queen Elizabeth II's relationships", 'the book is about Queen Elizabeth II', 'the book is set against the backdrop of royal duties'], 'dislike': ["The book ignores Queen Elizabeth II's relationships.", 'The book is not about Queen Elizabeth II.', 'The book is set against the backdrop of ordinary life.']}\}}& 9 & How much emphasis would you like the biography or memoir to place on historical context and the broader social and political climate of the time, versus a more intimate, personal focus on the individual's experiences? & I'm looking for a balance between personal experiences and historical context, especially how they dealt with challenges. & { \tiny \{{'preference': ['the book provides a balance between personal experiences and historical context', 'the book includes how the individual dealt with challenges', 'The author weaves together personal anecdotes with thorough historical research.', 'The author provides personal anecdotes about overcoming obstacles.', "The book strikes a balance between the author's personal stories and the broader historical background.", "The book discusses the individual's methods for overcoming obstacles."], 'dislike': ['The book lacks a balance between personal experiences and historical context.', 'The book excludes how the individual dealt with challenges.']}\}}& 39\\ \hline
\end{tabular}}
\caption{User and \system{} + Expansion interaction excerpts from Amazon book dataset. Context is provided to the user simulator, and rank is the current position of the target item among all candidates.}
\label{tab:interactions2}
\end{table*}

\end{document}